\documentclass[twocolumn,prl,superscriptaddress,longbibliography
]{revtex4-1}
\usepackage[english]{babel}
\usepackage{graphicx}
\usepackage{amssymb}
\usepackage{amsmath}
\usepackage{color}
\usepackage{comment}
\usepackage{bm}
\usepackage{babel}

\newcommand{\vect}[1]{{\mathbf #1}}
\renewcommand{\k}{{\bf k}}
\newcommand{\p}{{\bf p}}
\newcommand{\q}{{\bf q}}

\newcommand{\0}{{\bf 0}}

\newcommand{\am}{a_-}
\newcommand{\ab}{a_{\rm B}}

\newcommand{\bra}[1]{\langle\left.{#1}\right|}
\newcommand{\ket}[1]{\left|{#1}\right>}

\newcommand{\ek}{\epsilon_{\k}}

\newcommand{\nn}{\nonumber}
\newcommand{\beq}{\begin{equation}}
\newcommand{\eeq}{\end{equation}}

\newcommand{\sout}[1]{}

\begin{document}

\title{
Impurity in a Bose-Einstein condensate and the Efimov effect
  }

\author{Jesper Levinsen}
\affiliation{School of Physics and Astronomy, Monash University, Victoria 3800, Australia}

\author{Meera M.~Parish}
\affiliation{School of Physics and Astronomy, Monash University, Victoria 3800, Australia}
\affiliation{London Centre for Nanotechnology, Gordon Street, London, WC1H 0AH, United Kingdom}

\author{Georg M.~Bruun}
\affiliation{Department of Physics and Astronomy, Aarhus University, DK-8000 Aarhus C, Denmark}

\date{\today}

\begin{abstract}
We investigate the zero-temperature properties of an impurity
particle interacting with a Bose-Einstein condensate (BEC), using a
variational wavefunction that includes up to two Bogoliubov
excitations of the BEC. This allows one to capture three-body Efimov
physics, as well as to recover the first non-trivial terms in the
weak-coupling expansion.  We show that the energy and quasiparticle
residue of the dressed impurity (polaron) are significantly lowered
by three-body correlations, even for weak interactions where there
is no Efimov trimer state in a vacuum. For increasing attraction
between the impurity and the BEC, we observe a smooth crossover from
atom to Efimov trimer, with a superposition of states near the
Efimov resonance. We furthermore demonstrate that three-body loss
does not prohibit the experimental observation of these effects.
Our results thus suggest a route to realizing Efimov physics in a
stable quantum many-body system for the first time.
\end{abstract}


\maketitle

The underlying few-body dynamics in quantum many-body systems is
fundamental to our understanding of nature. For instance, Fermi liquid
theory is a spectacularly successful description of interacting
many-body systems in terms of the dynamics of single particles,
denoted
quasiparticles~\cite{Landau1957tof,BaymPethick1991book}. Likewise,
tightly bound two-body molecules in a molecular BEC are smoothly
connected to Cooper pairs in a Bardeen-Cooper-Schrieffer superfluid
with decreasing attraction, as has been beautifully demonstrated in
atomic gas experiments~\cite{Leggett1980,Nozieres1985,Randeria2014}.
It is thus natural to ask whether three-body bound states lead to any
observable effects in a many-body environment. In particular, Efimov
famously demonstrated that there are an infinite number of three-body
bound states (trimers) for identical bosons with a short range
resonant interaction characterized by an infinite scattering
length~\cite{Efimov1970,Braaten2006}. It took more than thirty years
before these Efimov states were first observed as a feature in the
three-body loss rate in non-degenerate atomic gases
\cite{Kraemer2006,zaccanti2009,Gross2009,Pollack2009,Lompe2010}, and
more recently through Coulomb explosion imaging of $^4$He trimers
\cite{Kunitski2015}. An intriguing question is whether Efimov states
can have observable effects on a quantum many-body system,
particularly outside the regime of strong three-body losses. So far,
there have been no experiments reporting such effects.

Here, we show how highly imbalanced two-component atomic gases can
provide a different, many-body signature of Efimov physics.  The study
of population imbalanced gases has already yielded fundamental new
insights concerning quasiparticles in Fermi systems (Fermi
polaron)~\cite{Schirotzek2009,Kohstall2012,Koschorreck2012,Massignan_Zaccanti_Bruun}. With
the recent identification of Feshbach resonances in Bose-Fermi and
Bose-Bose mixtures~\cite{Wu2012, Heo2012,Cumby2013,
  Roati2007,Pilch2009}, the study of quasiparticles in a BEC (Bose
polaron) with tunable interactions is now within reach.  Importantly,
there is no reason to expect three-body correlations to be small for
the Bose polaron, in contrast to the Fermi polaron, where these are
suppressed by Fermi statistics in the two-component
system~\cite{Combescot2008,Vlietinck2013}.  To have enhanced
three-body correlations in Fermi systems, one requires unequal
masses~\cite{Mathy2011}, low
dimensions~\cite{orso2010,Parish2011,Parish2013}, or at least three
different species~\cite{Nishida2015,cui2015}.

The large majority of theoretical studies of the Bose polaron have
been based on mean-field
theory~\cite{Astrakharchik2004,Cucchietti2006,Kalas2006,Bruderer2008,Volosniev2015},
or the Fr\"ohlich
model~\cite{Huang2009,Tempere2009,Casteels2014,Shashi2014,Grusdt2014,Shchadilova2014,Vlietinck2014}.
These approaches, however, do not include three-body Efimov
correlations, and the same applies to recent variational~\cite{Li2014}
and field theoretical~\cite{Rath2013} studies.  Only recently did a
perturbative calculation demonstrate that three-body correlations
indeed show up at third order in the impurity-boson scattering length
$a$, leading to important effects such as a logarithmic contribution
to the polaron energy~\cite{Christensen2015}.

In this Letter, we develop a variational ansatz for the Bose polaron
that includes three-body Efimov physics, and in addition recovers all
contributions to the energy in the weak-coupling limit up to and
including the logarithmic term at order $a^3$. We show that Efimov
correlations significantly lower the energy and quasiparticle residue,
even for scattering lengths substantially smaller in magnitude than
$\am<0$, the scattering length where the deepest Efimov trimer crosses
the three-body continuum in a vacuum. As $a$ approaches $\am$, the
energy exhibits a clear avoided crossing, reflecting a smooth change
in the quasiparticle character, from single-particle to three-body.
We furthermore demonstrate that these effects can be observed before
encountering resonantly enhanced three-body loss, which has so far
been the main experimental signature of Efimov physics.

\emph{Model.--} We consider an impurity particle immersed in a weakly
interacting BEC with $n_0\ab^3\ll 1$, where $n_0$ denotes the
condensate density and $\ab$ the boson-boson scattering length.  The
addition of an impurity atom to the condensate introduces two
additional length scales: The impurity-boson scattering length $a$,
and the so-called three-body parameter. The latter sets the scale of
one Efimov trimer in the universal limit of $1/a\to0$, from which all
other trimer energies can be determined. In the cold atoms context, it
is natural to use $\am$ as the three-body parameter.  It has recently
emerged in studies of identical
bosons~\cite{Berninger2011,Roy2013,Wang2012} that this parameter is
universally related to the van der Waals range.  Thus, we will fix
$\am$ through the two-body physics.

For this purpose, it is convenient to use the two-channel Hamiltonian
(setting $\hbar$ and the volume to 1):
\begin{align}\label{eq:Ham2}
  \hat{H} = & \sum_\k \left[ E_\k \beta^\dag_\k \beta_\k +
              \epsilon_\k
              c^\dag_\k c_\k + 
              \left( \epsilon^{\rm d}_\k + \nu_0 \right)
              d^\dag_\k d_\k \right ] \nn  \\ & \hspace{-10mm}
+g \sqrt{n_0} \sum_\k \left( d^\dag_\k c_\k+ h.c. \right)
+ g \sum_{\k,\q } \left(d^\dag_\q 
                                                           c_{\q -\k}
b_\k + h.c.   \right),
\end{align}
which measures the energy with respect to that of the BEC. Here,
$b^\dag_\k$ and $c^\dag_\k$ create a boson and the impurity,
respectively, with momentum $\k$ and single-particle energy
$\epsilon_\k = k^2/2m$. The impurity interacts with a boson by forming
a closed channel molecule, created by $d^\dag_\k$, with dispersion
$\epsilon_\k^{\rm d}= \ek/2$ and bare detuning $\nu_0$. The coupling
strength for this process, $g$, is taken constant up to a momentum
cut-off $\Lambda$.  We follow the usual weak-coupling prescription,
relating the boson operator $b$ to the Bogoliubov excitation $\beta$
via $b_\k=u_\k\beta_\k -v_\k\beta^\dag_{-\k}$ with real and positive
coherence factors $u_\k^2 = [1 + (\epsilon_\k + \mu)/E_\k]/2$ and
$v_\k^2 = [-1 + (\epsilon_\k + \mu)/E_\k]/2$, dispersion
$E_\k = \sqrt{\epsilon_\k (\epsilon_\k + 2\mu)}$, and chemical
potential $\mu = 4\pi \ab n_0/m$. The impurity is assumed to have the
same mass $m$ as the bosons, as we do not expect our results to change
qualitatively in the mass imbalanced system.

By considering impurity-boson scattering, we relate the bare molecule
detuning $\nu_0$, coupling $g$, and momentum cut-off to the scattering
length $a$ and the effective range $r_0$~\cite{Bruun2004,supmat}. The
additional length scale $r_0$ provided by the closed channel fixes the
three-body parameter of Efimov physics.  In particular, by solving the
impurity-boson-boson problem in vacuum with $\ab=0$, we find
$\am\approx 2467r_0$, which is consistent with calculations involving
realistic interatomic potentials for a range of heteronuclear
systems~\cite{Wang2012_2}.  This large separation of scales means we
can always take $|a/r_0|\gg1$, where the effect of the closed channel
on two-body processes is negligible.

\emph{Variational approach.--} 
We consider the wavefunction: 
\begin{align}
  \ket{\psi} = \,  \Bigg( &\alpha_0 c^\dag_{0} +  
                  \sum_\k \alpha_\k c^\dag_{- \k}
                  \beta^\dag_\k
+\frac{1}{2} \sum_{\k_1 \k_2} \alpha_{\k_1 \k_2} c^\dag_{- \k_1 - \k_2} 
                  \beta^\dag_{\k_1}   \beta^\dag_{\k_2} 
\nn \\ & +\gamma_0 d^\dag_0 + 
                  \sum_\k \gamma_\k d^\dag_{- \k} \beta^\dag_\k 
                  \Bigg) \ket{\Phi},
\label{eq:psi}
\end{align}
with $\ket{\Phi}$ the wavefunction of the interacting BEC. The first
line of Eq.~\eqref{eq:psi} describes the bare impurity and its
dressing by one or two Bogoliubov modes.  The second line describes
the impurity bound to a boson and forming a closed channel molecule,
which can be dressed by a Bogoliubov mode. The variational approach
consists in taking the stationary condition
$\bra{\partial\psi} (\hat{H} - E) \ket{\psi}=0$ where the derivative
is with respect to any of the expansion parameters. This procedure
yields five coupled equations for the energy $E$ \cite{supmat}.
Defining
\begin{align}
  \eta = & \ \frac{g^2 \sum_\k u_\k \alpha_\k}{E-\nu_0}, \hspace{5mm} %\nn \\
  \xi_\k =  \ \frac{g^2 \sum_\vect{k'} u_\vect{k'} \alpha_{\k\vect{k'}}}{E- \nu_0 -\epsilon^{d}_\k  - E_\k},
\end{align}
which remain finite in the limit $\nu_0 \to \infty$, and carrying out
the renormalization procedure, we finally arrive at the coupled
integral equations
\begin{gather} 
{\mathcal T}^{-1}(\0,E)
\eta = 
\frac{n_0}{E} \eta + \sqrt{n_0}
\sum_\k\left(\frac{u_\k\xi_\k}{E-\epsilon_\k-E_\k}
-\frac{v_\k \xi_\k}E \right), \nn
\\ \notag
 {\mathcal T}^{-1}(\k_1,E-E_{\k_1})\xi_{\k_1}   = 
\sqrt{n_0}\left(\frac{u_{\k_1}\eta}{E - \epsilon_{\k_1} 
-E_{\k_1}}-\frac{v_{\k_1}\eta}E\right) \nn \\ 
 +\frac{n_0\xi_{\k_1}}{E-\epsilon_{\k_1}-E_{\k_1}}+
\sum_{\k_2} \left(\frac{u_{\k_1} u_{\k_2}\xi_{\k_2}}{E_{\k_1 \k_2}}
+\frac{v_{\k_1} v_{\k_2}\xi_{\k_2}}E\right),
\label{eq:inteqs}
\end{gather}
where $E_{\k_1\k_2}\equiv E-E_{\k_1}-E_{\k_2}-\epsilon_{\k_1+\k_2}$.
Here, the medium ${\mathcal T}$ matrix is
\begin{align}
{\mathcal T}^{-1}(\k_1,E)= & \ \frac{m_r}{2\pi a}
-\frac{m_r^2r_0}{2\pi}(E-\epsilon_{\k_1}^{\rm d}) -\Pi(\k_1,E),
\label{eq:Tmed}
\end{align}
with
$\Pi(\k_1,E)=\sum_{\k_2}[u_{\k_2}^2/(E-E_{\k_2}-\epsilon_{{\mathbf
    k}_1+{\mathbf k}_2})+2m_r/k_2^2]$
the pair propagator in the BEC and $m_r = m/2$ the reduced mass.

A few comments on our variational approach are in order.  First, had
the two equations \eqref{eq:inteqs} not been coupled, the first of
these would simply reproduce the mean-field polaron energy,
$E=4\pi n_0a/m$ (for a non-zero $r_0$ there is a negligible
correction~$\sim{\cal O}(n^2a^3r_0)$ as $r_0$ is by far the smallest
length scale). Second, in the limit $n_0\to 0$ the second equation
exactly yields the Efimov spectrum for the impurity and two identical,
non-interacting bosons. This explicitly demonstrates that our
variational ansatz contains the relevant three-body Efimov
correlations.  For a discussion of the three-body problem with an
effective range, we refer the reader to
Refs.~\cite{Petrov2004tbp,Levinsen2009ads,Petrov2012tfa}.

Furthermore, since our wavefunction includes all two-body excitations,
it recovers all second order perturbation terms in $a$. This is
because all relevant Feynman diagrams contain at most two concurrent
boson propagators (neglecting $n_0\ab^3$
corrections)~\cite{Christensen2015}. Only a subset of these diagrams
are produced if one restricts the approach to one Bogoliubov mode, as
in Ref.~\cite{Li2014}. Although our ansatz does not reproduce all
third order self energy diagrams, it does indeed capture the
logarithmic term in the energy, since this originates from diagrams
containing at most two simultaneous Bogoliubov
excitations~\cite{Christensen2015}.

Finally, we remark that had we used a single-channel model from the
outset, we would have arrived at equations identical to
Eqs.~\eqref{eq:inteqs} with $r_0=0$ in Eq.~\eqref{eq:Tmed}. However,
in this case we would need to set the three-body parameter in some
other way, for instance through the use of a Gaussian cut-off.

%%%%%%%%%%%%%%%%%%%%%%%%%%%%%%%%%%%%%%%%%%%%%%%%%%%%
\begin{figure}[t]
\begin{center}
\includegraphics[width=.85\columnwidth]{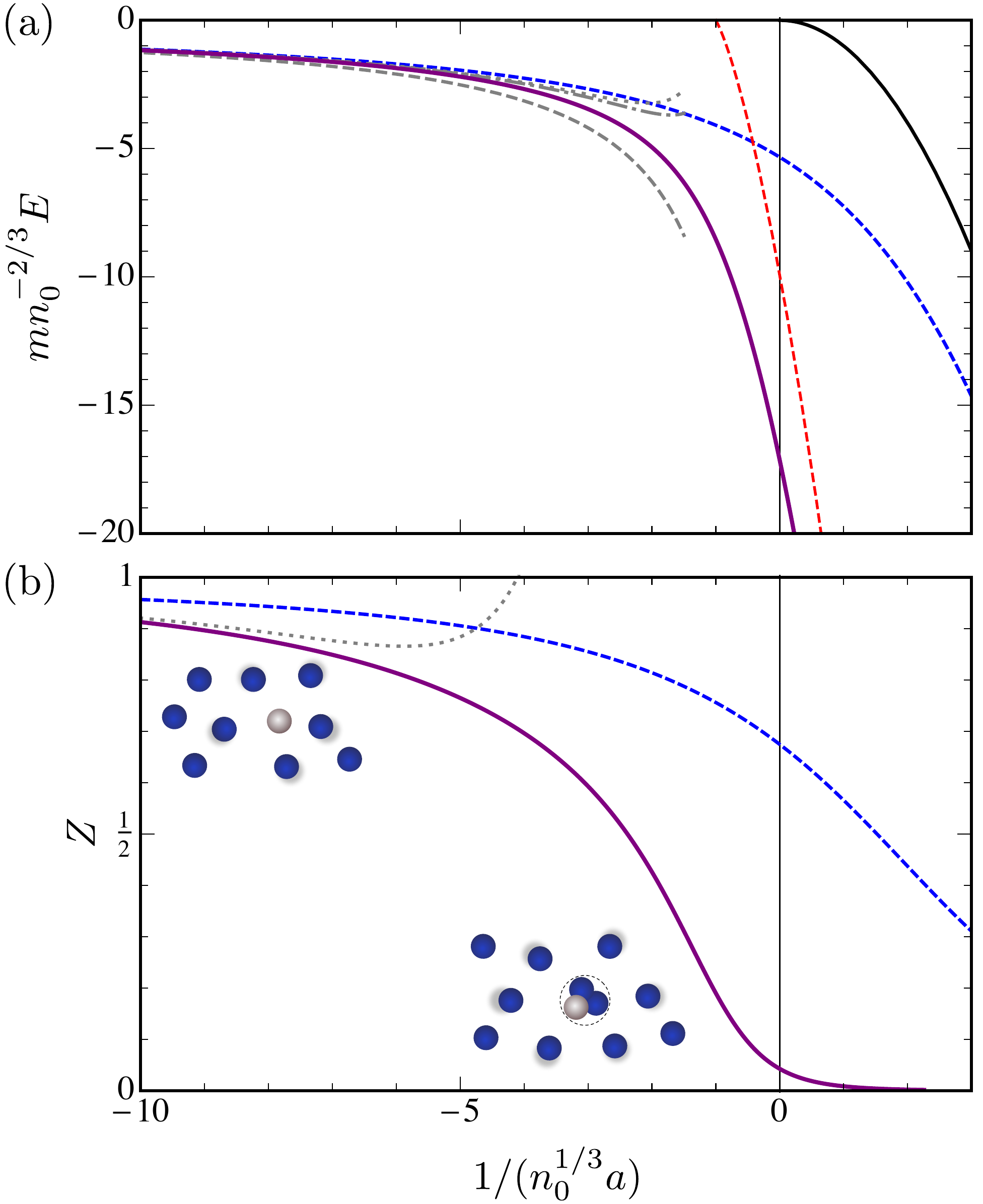}
\caption{(a) Spectrum and (b) residue of the Bose polaron for
  $n_0^{1/3}\am=-1$ and $\ab/\am=-1/50$. We show the results from our
  variational calculation (purple, solid), and the variational
  wavefunction excluding three-body correlations and effective range
  effects (blue, dashed). The gray lines are the results of
  perturbation theory: Mean-field (dashed), second
  order~\cite{Casteels2014} (dot-dashed), and third
  order~\cite{Christensen2015} (dotted). We also display the energy of
  the deepest Efimov state in a vacuum (red, dashed), and for $a>0$
  the threshold of the atom-dimer continuum given by the
  impurity-boson bound state energy
  $-(\sqrt{1-2r_0/a}-1)^2/(2m_rr_0^2)$ (black, solid). The insets
  illustrate the crossover in the wavefunction (the atoms are not
  drawn to scale). }
\label{fig:EZam1}
\end{center}
\end{figure}
%%%%%%%%%%%%%%%%%%%%%%%%%%%%%%%%%%%%%%%%%%%%%%%%%%%%

\emph{Results.--} The Bose polaron is characterized by its
quasiparticle properties, i.e.~the energy $E$ and residue
$Z\equiv|\alpha_0|^2$~\cite{supmat}.  In Fig.\ \ref{fig:EZam1}, we
display these quantities calculated from Eq.~\eqref{eq:inteqs} as a
function of $1/(n_0^{1/3}a)$ for realistic experimental parameters.
For weak interaction, $1/(n_0^{1/3}a)\ll -1$, the polaron energy is
close to the mean-field value $4\pi an_0/m$ and $Z\simeq 1$.  With
increasing coupling, $E$ starts to deviate from the mean-field result,
whereas there is better agreement with second \cite{Casteels2014} and
third \cite{Christensen2015} order perturbation theory. This is as
expected since, as argued above, our variational approach contains all
diagrams contributing to the energy at these orders of perturbation
theory. While $Z=1$ within mean-field theory, and the second order
perturbative expression~\cite{Christensen2015} is negative in the
range plotted, the agreement at third order with our variational
ansatz is very good, leading us to conclude that although the
variational ansatz does not contain all third order terms for the
residue, it likely includes the most important ones.

%%%%%%%%%%%%%%%%%%%%%%%%%%%%%%%%%%%%%%%%%%%%%%%%%%%%
\begin{figure}[t]
\begin{center}
 \includegraphics[width=0.95\columnwidth]{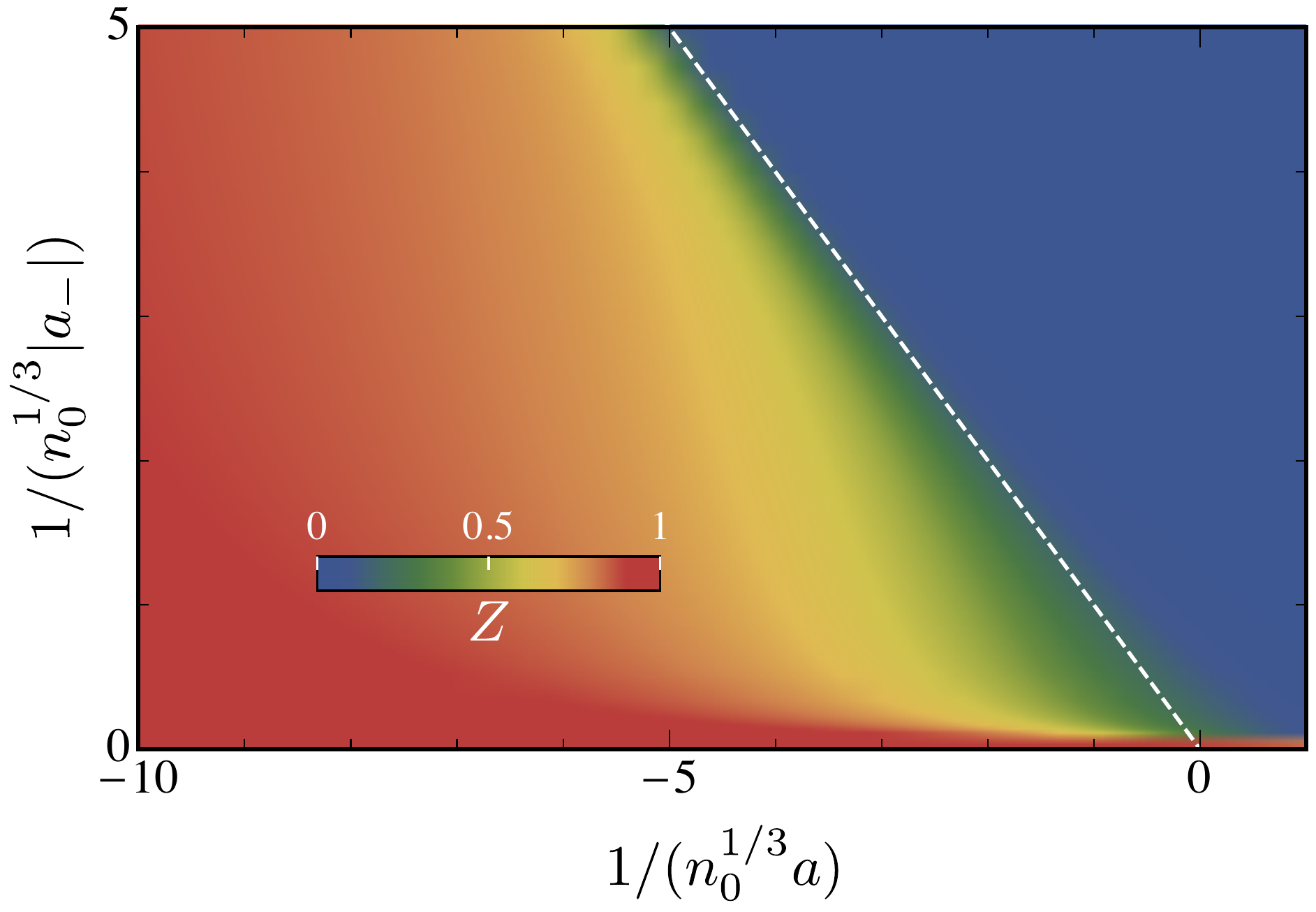}
 \caption{Residue of the Bose polaron as a function of
   $1/(n_0^{1/3}a)$ and $1/(n_0^{1/3}|\am|)$ for $\ab/\am=-1/50$.  The
   white dashed line is $a=\am$.  For sufficiently small
   $1/(n_0^{1/3}|\am|)$, the residue remains close to 1 around
   unitarity, implying that the Efimov trimer has disappeared.}
\label{fig:densplot}
\end{center}
\end{figure}
%%%%%%%%%%%%%%%%%%%%%%%%%%%%%%%%%%%%%%%%%%%%%%%%%%%%
 
Closer to resonance, three-body correlations become important, leading
to an avoided crossing between a state best described as a single
particle interacting attractively with the medium and an Efimov trimer
dressed by interactions, as seen in Fig.\ \ref{fig:EZam1} and
illustrated in the insets. The residue correspondingly decreases as
the wavefunction resembles the three-body Efimov state. This crossover
is of course completely missed by the perturbative expressions. It is
also missed by the variational ansatz used in Ref.~\cite{Li2014},
which includes only two-body correlations, i.e., only the first two
terms in Eq.~\eqref{eq:psi}. Within this approach the quasiparticle
residue only goes to zero once the polaron approaches the atom-dimer
continuum.  Our analysis is the first to include Efimov physics for
the Bose polaron, which we see strongly changes its properties at a
qualitative level.  Even outside the regime of Efimov physics in
vacuum, there are strong effects of Efimov physics: For instance, at
$a=\am/2$ the magnitude of the energy is increased by $50\%$ when
comparing with the variational approach restricted to two-body
correlations.

To investigate the strength of the coupling between the two branches
involved in the avoided crossing, we show in Fig.~\ref{fig:densplot}
the residue as a function of $1/(n_0^{1/3}a)$ and
$1/(n_0^{1/3}|\am|)$. Focussing first on the region where
$1/(n_0^{1/3}|\am|)\geq1$, we see that the crossover region is located
around $a\sim\am$, where $Z$ decreases to almost zero. Furthermore,
the range of scattering lengths over which this transition occurs
decreases with decreasing $|\am|$, reflecting how the coupling to the
Efimov state becomes weaker. This can be understood from a simple
dimensional analysis as follows: Treating momentarily the Efimov
trimer as a point particle, the coupling of this state to two bosons
and the impurity involves four particles, three in-going and one
out-going or vice versa. The coupling strength for this process must
therefore scale as Length/Mass. In the avoided crossing region, there
is only one relevant length scale, $a\sim\am$, and consequently the
coupling strength $\sim \am/m$. On the other hand, in the regime
$1/(n_0^{1/3}|\am|)\ll1$ the perturbative energy far exceeds that of
the Efimov trimer in vacuum when $a\sim\am$ and thus the quasiparticle
remains well-defined even beyond this point, as seen in the
figure. Eventually, of course, the residue is suppressed as the
polaron becomes dimer-like for sufficiently small and positive $a$.

\begin{figure}[t]
\begin{center}
\includegraphics[width=0.85\columnwidth]{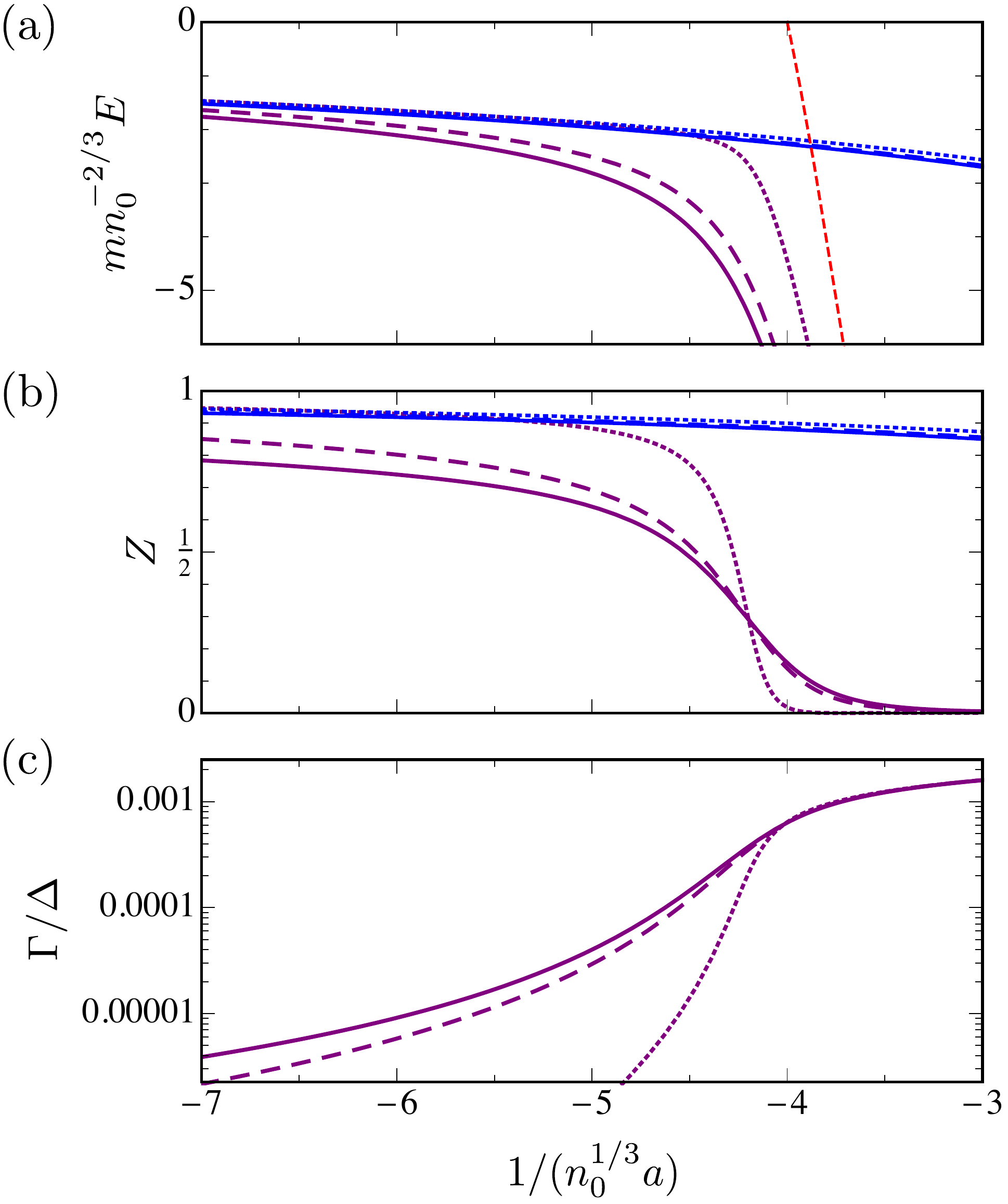}
\caption{(a) Energy, (b) residue, and (c) three-body loss rate for
  $n_0^{1/3}|\am|=0.25$ and different values of $\ab/|\am|$: 0.01
  (purple, solid), 0.1 (purple, dashed), 0.5 (purple, dotted). In (a)
  we also display the energy of the deepest Efimov state in vacuum
  (red, dashed).  The blue lines are the variational results excluding
  three-body correlations as well as effective range effects.}
\label{fig:ab}
\end{center}
\end{figure}

To investigate the dependence on $\ab$, we plot in Fig.\
\ref{fig:ab}(a-b) the energy and residue of the Bose polaron for
$n_0^{1/3}\am=-1/4$ and three different values of the boson-boson
scattering length. In both quantities we clearly observe how the
crossover becomes sharper with increasing $\ab/|\am|$, since
boson-boson repulsion suppresses the formation of the Efimov trimer.
An even larger suppression is expected to occur for any four-body
Efimov state \cite{Stecher2008} consisting of the impurity and three
bosons, if such states exist (these have so far only been investigated
for unequal masses \cite{Blume2014}).  Indeed, the impact of four-body
states on the many-body system will be further reduced since one
expects them to appear for $|a|<|\am|$, resulting in a smaller
coupling to the bound state similar to that shown in
Fig.~\ref{fig:densplot}. It is thus likely that such states can be
ignored even in a regime where the coupling to the three-body state is
strong.

While the energy and residue of the Fermi polaron have been measured
using radio-frequency (RF)
spectroscopy~\cite{Schirotzek2009,Kohstall2012,Koschorreck2012}, the
presence of the Efimov trimer in a cold atoms context has hitherto
only been observed indirectly, as either a significant increase
(suppression) in the three-body loss rate for negative (positive)
scattering lengths
\cite{Kraemer2006,zaccanti2009,Gross2009,Pollack2009}, or via enhanced
loss from the RF association of trimers close to the atom-dimer
threshold \cite{Lompe2010}. An important question is therefore whether
the shifts in the polaron energy and residue due to Efimov
correlations can be observed before encountering prohibitively fast
losses. To investigate this, we note that the main loss process --
three-body recombination to a deeply bound molecule and an atom -- is
local, since it requires three atoms to be separated within a distance
comparable to the size of the outgoing states. Hence, the rate of this
loss is captured, within a perturbative approach, by the expectation
value~\cite{Braaten2001,Levinsen2011}
\begin{align}
\Gamma=\Delta\langle\psi|\sum_{\k_1\k_2\p}d_{\p-\k_1}^\dagger b_{\k_1}^\dagger b_{\k_2} d_{\p-\k_2}|\psi\rangle.
\label{eq:Hloss}
\end{align}
Making the replacement $\beta_k\to b_k$ at short range, we get
$\Gamma\approx\Delta \sum_k|\gamma_k|^2$.  Note that the constant
$\Delta$ should be fixed by comparison with experimental data outside
the regime of resonantly enhanced losses; thus, in Fig.\
\ref{fig:ab}(c) we plot $\Gamma/\Delta$. We see that the loss rate
increases with decreasing $Z$ due to the onset of Efimov correlations,
and saturates to that of the vacuum Efimov state. Figure \ref{fig:ab},
however, demonstrates a crucial point: The reduction in the energy and
residue due to three-body correlations is still substantial away from
$a=\am$, where the decay rate can be more than \emph{two orders of
  magnitude} smaller than that of the vacuum Efimov state. From this
we conclude that the lifetime of the impurity is indeed long enough in
this regime to allow a detection of three-body effects.  This mirrors
previous work on trimers in a quasi-two-dimensional geometry, which
argues that Efimov physics can be probed by coupling an Efimov trimer
to a trimer extended within the plane~\cite{Levinsen2014}.

 %%%%%%%%%%%%%%%%%%%%%%%%%%%%%%%%%%%%%%%%%%%%%%%%%%%%
%%%%%%%%%%%%%%%%%%%%%%%%%%%%%%%%%%%%%%%%%%%%%%%%%%%%

\emph{Conclusions.--} Using a variational ansatz which includes
three-body Efimov correlations as well as the lowest non-trivial terms
in the weak-coupling expansion, we demonstrated that Efimov physics
dramatically changes the properties of the Bose polaron. The energy
and quasiparticle residue are lowered significantly, even when there
is no Efimov trimer state in a vacuum. As the critical scattering
length for the emergence of the trimer state is approached, the energy
exhibits an avoided crossing and the residue quickly decreases,
reflecting the fact that the impurity smoothly becomes bound in an
Efimov trimer. Crucially, we showed that three-body loss associated
with the Efimov state does not prohibit the observation of these
effects. In conclusion, our results demonstrate how Efimov physics
qualitatively changes the properties of a quantum many-body system,
and how it can be observed in signals other than
loss.

\begin{acknowledgments}
  We gratefully acknowledge fruitful discussion with Jan Arlt, Rasmus
  S.\ Christensen, Yusuke Nishida, and Richard Schmidt.  G.M.B.~wishes
  to acknowledge the support of the Villum Foundation via grant
  VKR023163.
\end{acknowledgments}

\bibliography{bosepolaron}

\begin{thebibliography}{59}%
\makeatletter
\providecommand \@ifxundefined [1]{%
 \@ifx{#1\undefined}
}%
\providecommand \@ifnum [1]{%
 \ifnum #1\expandafter \@firstoftwo
 \else \expandafter \@secondoftwo
 \fi
}%
\providecommand \@ifx [1]{%
 \ifx #1\expandafter \@firstoftwo
 \else \expandafter \@secondoftwo
 \fi
}%
\providecommand \natexlab [1]{#1}%
\providecommand \enquote  [1]{``#1''}%
\providecommand \bibnamefont  [1]{#1}%
\providecommand \bibfnamefont [1]{#1}%
\providecommand \citenamefont [1]{#1}%
\providecommand \href@noop [0]{\@secondoftwo}%
\providecommand \href [0]{\begingroup \@sanitize@url \@href}%
\providecommand \@href[1]{\@@startlink{#1}\@@href}%
\providecommand \@@href[1]{\endgroup#1\@@endlink}%
\providecommand \@sanitize@url [0]{\catcode `\\12\catcode `\$12\catcode
  `\&12\catcode `\#12\catcode `\^12\catcode `\_12\catcode `\%12\relax}%
\providecommand \@@startlink[1]{}%
\providecommand \@@endlink[0]{}%
\providecommand \url  [0]{\begingroup\@sanitize@url \@url }%
\providecommand \@url [1]{\endgroup\@href {#1}{\urlprefix }}%
\providecommand \urlprefix  [0]{URL }%
\providecommand \Eprint [0]{\href }%
\providecommand \doibase [0]{http://dx.doi.org/}%
\providecommand \selectlanguage [0]{\@gobble}%
\providecommand \bibinfo  [0]{\@secondoftwo}%
\providecommand \bibfield  [0]{\@secondoftwo}%
\providecommand \translation [1]{[#1]}%
\providecommand \BibitemOpen [0]{}%
\providecommand \bibitemStop [0]{}%
\providecommand \bibitemNoStop [0]{.\EOS\space}%
\providecommand \EOS [0]{\spacefactor3000\relax}%
\providecommand \BibitemShut  [1]{\csname bibitem#1\endcsname}%
\let\auto@bib@innerbib\@empty
%</preamble>
\bibitem [{\citenamefont {Landau}(1957)}]{Landau1957tof}%
  \BibitemOpen
  \bibfield  {author} {\bibinfo {author} {\bibfnamefont {L.}~\bibnamefont
  {Landau}},\ }\bibfield  {title} {\bibinfo {title} {\emph {Theory of
  Fermi-liquids}},\ }\href@noop {} {\bibfield  {journal} {\bibinfo  {journal}
  {Sov. Phys. JETP}\ }\textbf {\bibinfo {volume} {3}},\ \bibinfo {pages} {920}
  (\bibinfo {year} {1957})}\BibitemShut {NoStop}%
\bibitem [{\citenamefont {Baym}\ and\ \citenamefont
  {Pethick}(1991)}]{BaymPethick1991book}%
  \BibitemOpen
  \bibfield  {author} {\bibinfo {author} {\bibfnamefont {G.}~\bibnamefont
  {Baym}}\ and\ \bibinfo {author} {\bibfnamefont {C.}~\bibnamefont {Pethick}},\
  }\href@noop {} {\emph {\bibinfo {title} {Landau Fermi-Liquid Theory: Concepts
  and Applications}}}\ (\bibinfo  {publisher} {Wiley-VCH},\ \bibinfo {year}
  {1991})\BibitemShut {NoStop}%
\bibitem [{\citenamefont {Leggett}(1980)}]{Leggett1980}%
  \BibitemOpen
  \bibfield  {author} {\bibinfo {author} {\bibfnamefont {A.}~\bibnamefont
  {Leggett}},\ }\bibfield  {title} {\bibinfo {title} {\emph {Diatomic molecules
  and cooper pairs}},\ }in\ \href {\doibase 10.1007/BFb0120125} {\emph
  {\bibinfo {booktitle} {Modern Trends in the Theory of Condensed Matter}}},\
  \bibinfo {series} {Lecture Notes in Physics}, Vol.\ \bibinfo {volume} {115},\
  \bibinfo {editor} {edited by\ \bibinfo {editor} {\bibfnamefont
  {A.}~\bibnamefont {Pƒ{\^o}kalski}}\ and\ \bibinfo {editor} {\bibfnamefont
  {J.}~\bibnamefont {Przystawa}}}\ (\bibinfo  {publisher} {Springer Berlin
  Heidelberg},\ \bibinfo {year} {1980})\ pp.\ \bibinfo {pages}
  {13--27}\BibitemShut {NoStop}%
\bibitem [{\citenamefont {Nozi\'eres}\ and\ \citenamefont
  {Schmitt-Rink}(1985)}]{Nozieres1985}%
  \BibitemOpen
  \bibfield  {author} {\bibinfo {author} {\bibfnamefont {P.}~\bibnamefont
  {Nozi\'eres}}\ and\ \bibinfo {author} {\bibfnamefont {S.}~\bibnamefont
  {Schmitt-Rink}},\ }\bibfield  {title} {\bibinfo {title} {\emph {Bose
  condensation in an attractive fermion gas: From weak to strong coupling
  superconductivity}},\ }\href {\doibase 10.1007/BF00683774} {\bibfield
  {journal} {\bibinfo  {journal} {Journal of Low Temperature Physics}\ }\textbf
  {\bibinfo {volume} {59}},\ \bibinfo {pages} {195} (\bibinfo {year}
  {1985})}\BibitemShut {NoStop}%
\bibitem [{\citenamefont {Randeria}\ and\ \citenamefont
  {Taylor}(2014)}]{Randeria2014}%
  \BibitemOpen
  \bibfield  {author} {\bibinfo {author} {\bibfnamefont {M.}~\bibnamefont
  {Randeria}}\ and\ \bibinfo {author} {\bibfnamefont {E.}~\bibnamefont
  {Taylor}},\ }\bibfield  {title} {\bibinfo {title} {\emph {Crossover from
  Bardeen-Cooper-Schrieffer to Bose-Einstein Condensation and the Unitary Fermi
  Gas}},\ }\href {\doibase 10.1146/annurev-conmatphys-031113-133829} {\bibfield
   {journal} {\bibinfo  {journal} {Annual Review of Condensed Matter Physics}\
  }\textbf {\bibinfo {volume} {5}},\ \bibinfo {pages} {209} (\bibinfo {year}
  {2014})}\BibitemShut {NoStop}%
\bibitem [{\citenamefont {Efimov}(1970)}]{Efimov1970}%
  \BibitemOpen
  \bibfield  {author} {\bibinfo {author} {\bibfnamefont {V.}~\bibnamefont
  {Efimov}},\ }\bibfield  {title} {\bibinfo {title} {\emph {Energy levels
  arising from resonant two-body forces in a three-body system}},\ }\href@noop
  {} {\bibfield  {journal} {\bibinfo  {journal} {Phys. Lett. B}\ }\textbf
  {\bibinfo {volume} {33}},\ \bibinfo {pages} {563} (\bibinfo {year}
  {1970})}\BibitemShut {NoStop}%
\bibitem [{\citenamefont {Braaten}\ and\ \citenamefont
  {Hammer}(2006)}]{Braaten2006}%
  \BibitemOpen
  \bibfield  {author} {\bibinfo {author} {\bibfnamefont {E.}~\bibnamefont
  {Braaten}}\ and\ \bibinfo {author} {\bibfnamefont {H.-W.}\ \bibnamefont
  {Hammer}},\ }\bibfield  {title} {\bibinfo {title} {\emph {{Universality in
  few-body systems with large scattering length}}},\ }\href@noop {} {\bibfield
  {journal} {\bibinfo  {journal} {Physics Reports}\ }\textbf {\bibinfo {volume}
  {428}},\ \bibinfo {pages} {259} (\bibinfo {year} {2006})}\BibitemShut
  {NoStop}%
\bibitem [{\citenamefont {{Kraemer}}\ \emph {et~al.}(2006)\citenamefont
  {{Kraemer}}, \citenamefont {{Mark}}, \citenamefont {{Waldburger}},
  \citenamefont {{Danzl}}, \citenamefont {{Chin}}, \citenamefont {{Engeser}},
  \citenamefont {{Lange}}, \citenamefont {{Pilch}}, \citenamefont {{Jaakkola}},
  \citenamefont {{N{\"a}gerl}},\ and\ \citenamefont {{Grimm}}}]{Kraemer2006}%
  \BibitemOpen
  \bibfield  {author} {\bibinfo {author} {\bibfnamefont {T.}~\bibnamefont
  {{Kraemer}}}, \bibinfo {author} {\bibfnamefont {M.}~\bibnamefont {{Mark}}},
  \bibinfo {author} {\bibfnamefont {P.}~\bibnamefont {{Waldburger}}}, \bibinfo
  {author} {\bibfnamefont {J.~G.}\ \bibnamefont {{Danzl}}}, \bibinfo {author}
  {\bibfnamefont {C.}~\bibnamefont {{Chin}}}, \bibinfo {author} {\bibfnamefont
  {B.}~\bibnamefont {{Engeser}}}, \bibinfo {author} {\bibfnamefont {A.~D.}\
  \bibnamefont {{Lange}}}, \bibinfo {author} {\bibfnamefont {K.}~\bibnamefont
  {{Pilch}}}, \bibinfo {author} {\bibfnamefont {A.}~\bibnamefont {{Jaakkola}}},
  \bibinfo {author} {\bibfnamefont {H.-C.}\ \bibnamefont {{N{\"a}gerl}}}, \
  and\ \bibinfo {author} {\bibfnamefont {R.}~\bibnamefont {{Grimm}}},\
  }\bibfield  {title} {\bibinfo {title} {\emph {{Evidence for Efimov quantum
  states in an ultracold gas of caesium atoms}}},\ }\href@noop {} {\bibfield
  {journal} {\bibinfo  {journal} {Nature}\ }\textbf {\bibinfo {volume} {440}},\
  \bibinfo {pages} {315} (\bibinfo {year} {2006})}\BibitemShut {NoStop}%
\bibitem [{\citenamefont {Zaccanti}\ \emph {et~al.}(2009)\citenamefont
  {Zaccanti}, \citenamefont {Deissler}, \citenamefont {D'Errico}, \citenamefont
  {Fattori}, \citenamefont {Jona-Lasinio}, \citenamefont {Muller},
  \citenamefont {Roati}, \citenamefont {Inguscio},\ and\ \citenamefont
  {Modugno}}]{zaccanti2009}%
  \BibitemOpen
  \bibfield  {author} {\bibinfo {author} {\bibfnamefont {M.}~\bibnamefont
  {Zaccanti}}, \bibinfo {author} {\bibfnamefont {B.}~\bibnamefont {Deissler}},
  \bibinfo {author} {\bibfnamefont {C.}~\bibnamefont {D'Errico}}, \bibinfo
  {author} {\bibfnamefont {M.}~\bibnamefont {Fattori}}, \bibinfo {author}
  {\bibfnamefont {M.}~\bibnamefont {Jona-Lasinio}}, \bibinfo {author}
  {\bibfnamefont {S.}~\bibnamefont {Muller}}, \bibinfo {author} {\bibfnamefont
  {G.}~\bibnamefont {Roati}}, \bibinfo {author} {\bibfnamefont
  {M.}~\bibnamefont {Inguscio}}, \ and\ \bibinfo {author} {\bibfnamefont
  {G.}~\bibnamefont {Modugno}},\ }\bibfield  {title} {\bibinfo {title} {\emph
  {Observation of an {Efimov} spectrum in an atomic system}},\ }\href@noop {}
  {\bibfield  {journal} {\bibinfo  {journal} {Nature Phys.}\ }\textbf {\bibinfo
  {volume} {5}},\ \bibinfo {pages} {586} (\bibinfo {year} {2009})}\BibitemShut
  {NoStop}%
\bibitem [{\citenamefont {Gross}\ \emph {et~al.}(2009)\citenamefont {Gross},
  \citenamefont {Shotan}, \citenamefont {Kokkelmans},\ and\ \citenamefont
  {Khaykovich}}]{Gross2009}%
  \BibitemOpen
  \bibfield  {author} {\bibinfo {author} {\bibfnamefont {N.}~\bibnamefont
  {Gross}}, \bibinfo {author} {\bibfnamefont {Z.}~\bibnamefont {Shotan}},
  \bibinfo {author} {\bibfnamefont {S.}~\bibnamefont {Kokkelmans}}, \ and\
  \bibinfo {author} {\bibfnamefont {L.}~\bibnamefont {Khaykovich}},\ }\bibfield
   {title} {\bibinfo {title} {\emph {Observation of Universality in Ultracold
  $^{7}\mathrm{Li}$ Three-Body Recombination}},\ }\href@noop {} {\bibfield
  {journal} {\bibinfo  {journal} {Phys. Rev. Lett.}\ }\textbf {\bibinfo
  {volume} {103}},\ \bibinfo {pages} {163202} (\bibinfo {year}
  {2009})}\BibitemShut {NoStop}%
\bibitem [{\citenamefont {Pollack}\ \emph {et~al.}(2009)\citenamefont
  {Pollack}, \citenamefont {Dries},\ and\ \citenamefont {Hulet}}]{Pollack2009}%
  \BibitemOpen
  \bibfield  {author} {\bibinfo {author} {\bibfnamefont {S.~E.}\ \bibnamefont
  {Pollack}}, \bibinfo {author} {\bibfnamefont {D.}~\bibnamefont {Dries}}, \
  and\ \bibinfo {author} {\bibfnamefont {R.~G.}\ \bibnamefont {Hulet}},\
  }\bibfield  {title} {\bibinfo {title} {\emph {Universality in Three- and
  Four-Body Bound States of Ultracold Atoms}},\ }\href@noop {} {\bibfield
  {journal} {\bibinfo  {journal} {Science}\ }\textbf {\bibinfo {volume}
  {326}},\ \bibinfo {pages} {1683} (\bibinfo {year} {2009})}\BibitemShut
  {NoStop}%
\bibitem [{\citenamefont {Lompe}\ \emph {et~al.}(2010)\citenamefont {Lompe},
  \citenamefont {Ottenstein}, \citenamefont {Serwane}, \citenamefont {Wenz},
  \citenamefont {Z{\"u}rn},\ and\ \citenamefont {Jochim}}]{Lompe2010}%
  \BibitemOpen
  \bibfield  {author} {\bibinfo {author} {\bibfnamefont {T.}~\bibnamefont
  {Lompe}}, \bibinfo {author} {\bibfnamefont {T.~B.}\ \bibnamefont
  {Ottenstein}}, \bibinfo {author} {\bibfnamefont {F.}~\bibnamefont {Serwane}},
  \bibinfo {author} {\bibfnamefont {A.~N.}\ \bibnamefont {Wenz}}, \bibinfo
  {author} {\bibfnamefont {G.}~\bibnamefont {Z{\"u}rn}}, \ and\ \bibinfo
  {author} {\bibfnamefont {S.}~\bibnamefont {Jochim}},\ }\bibfield  {title}
  {\bibinfo {title} {\emph {Radio-Frequency Association of {Efimov} Trimers}},\
  }\href@noop {} {\bibfield  {journal} {\bibinfo  {journal} {Science}\ }\textbf
  {\bibinfo {volume} {330}},\ \bibinfo {pages} {940} (\bibinfo {year}
  {2010})}\BibitemShut {NoStop}%
\bibitem [{\citenamefont {Kunitski}\ \emph {et~al.}(2015)\citenamefont
  {Kunitski}, \citenamefont {Zeller}, \citenamefont {Voigtsberger},
  \citenamefont {Kalinin}, \citenamefont {Schmidt}, \citenamefont
  {Sch{\"o}ffler}, \citenamefont {Czasch}, \citenamefont {Sch{\"o}llkopf},
  \citenamefont {Grisenti}, \citenamefont {Jahnke}, \citenamefont {Blume},\
  and\ \citenamefont {D{\"o}rner}}]{Kunitski2015}%
  \BibitemOpen
  \bibfield  {author} {\bibinfo {author} {\bibfnamefont {M.}~\bibnamefont
  {Kunitski}}, \bibinfo {author} {\bibfnamefont {S.}~\bibnamefont {Zeller}},
  \bibinfo {author} {\bibfnamefont {J.}~\bibnamefont {Voigtsberger}}, \bibinfo
  {author} {\bibfnamefont {A.}~\bibnamefont {Kalinin}}, \bibinfo {author}
  {\bibfnamefont {L.~P.~H.}\ \bibnamefont {Schmidt}}, \bibinfo {author}
  {\bibfnamefont {M.}~\bibnamefont {Sch{\"o}ffler}}, \bibinfo {author}
  {\bibfnamefont {A.}~\bibnamefont {Czasch}}, \bibinfo {author} {\bibfnamefont
  {W.}~\bibnamefont {Sch{\"o}llkopf}}, \bibinfo {author} {\bibfnamefont
  {R.~E.}\ \bibnamefont {Grisenti}}, \bibinfo {author} {\bibfnamefont
  {T.}~\bibnamefont {Jahnke}}, \bibinfo {author} {\bibfnamefont
  {D.}~\bibnamefont {Blume}}, \ and\ \bibinfo {author} {\bibfnamefont
  {R.}~\bibnamefont {D{\"o}rner}},\ }\bibfield  {title} {\bibinfo {title}
  {\emph {Observation of the Efimov state of the helium trimer}},\ }\href
  {\doibase 10.1126/science.aaa5601} {\bibfield  {journal} {\bibinfo  {journal}
  {Science}\ }\textbf {\bibinfo {volume} {348}},\ \bibinfo {pages} {551}
  (\bibinfo {year} {2015})}\BibitemShut {NoStop}%
\bibitem [{\citenamefont {Schirotzek}\ \emph {et~al.}(2009)\citenamefont
  {Schirotzek}, \citenamefont {Wu}, \citenamefont {Sommer},\ and\ \citenamefont
  {Zwierlein}}]{Schirotzek2009}%
  \BibitemOpen
  \bibfield  {author} {\bibinfo {author} {\bibfnamefont {A.}~\bibnamefont
  {Schirotzek}}, \bibinfo {author} {\bibfnamefont {C.-H.}\ \bibnamefont {Wu}},
  \bibinfo {author} {\bibfnamefont {A.}~\bibnamefont {Sommer}}, \ and\ \bibinfo
  {author} {\bibfnamefont {M.~W.}\ \bibnamefont {Zwierlein}},\ }\bibfield
  {title} {\bibinfo {title} {\emph {Observation of Fermi Polarons in a Tunable
  Fermi Liquid of Ultracold Atoms}},\ }\href {\doibase
  10.1103/PhysRevLett.102.230402} {\bibfield  {journal} {\bibinfo  {journal}
  {Phys. Rev. Lett.}\ }\textbf {\bibinfo {volume} {102}},\ \bibinfo {pages}
  {230402} (\bibinfo {year} {2009})}\BibitemShut {NoStop}%
\bibitem [{\citenamefont {{Kohstall}}\ \emph {et~al.}(2012)\citenamefont
  {{Kohstall}}, \citenamefont {{Zaccanti}}, \citenamefont {{Jag}},
  \citenamefont {{Trenkwalder}}, \citenamefont {{Massignan}}, \citenamefont
  {{Bruun}}, \citenamefont {{Schreck}},\ and\ \citenamefont
  {{Grimm}}}]{Kohstall2012}%
  \BibitemOpen
  \bibfield  {author} {\bibinfo {author} {\bibfnamefont {C.}~\bibnamefont
  {{Kohstall}}}, \bibinfo {author} {\bibfnamefont {M.}~\bibnamefont
  {{Zaccanti}}}, \bibinfo {author} {\bibfnamefont {M.}~\bibnamefont {{Jag}}},
  \bibinfo {author} {\bibfnamefont {A.}~\bibnamefont {{Trenkwalder}}}, \bibinfo
  {author} {\bibfnamefont {P.}~\bibnamefont {{Massignan}}}, \bibinfo {author}
  {\bibfnamefont {G.~M.}\ \bibnamefont {{Bruun}}}, \bibinfo {author}
  {\bibfnamefont {F.}~\bibnamefont {{Schreck}}}, \ and\ \bibinfo {author}
  {\bibfnamefont {R.}~\bibnamefont {{Grimm}}},\ }\bibfield  {title} {\bibinfo
  {title} {\emph {{Metastability and coherence of repulsive polarons in a
  strongly interacting Fermi mixture}}},\ }\href {\doibase 10.1038/nature11065}
  {\bibfield  {journal} {\bibinfo  {journal} {\nat}\ }\textbf {\bibinfo
  {volume} {485}},\ \bibinfo {pages} {615} (\bibinfo {year}
  {2012})}\BibitemShut {NoStop}%
\bibitem [{\citenamefont {{Koschorreck}}\ \emph {et~al.}(2012)\citenamefont
  {{Koschorreck}}, \citenamefont {{Pertot}}, \citenamefont {{Vogt}},
  \citenamefont {{Fr{\"o}hlich}}, \citenamefont {{Feld}},\ and\ \citenamefont
  {{K{\"o}hl}}}]{Koschorreck2012}%
  \BibitemOpen
  \bibfield  {author} {\bibinfo {author} {\bibfnamefont {M.}~\bibnamefont
  {{Koschorreck}}}, \bibinfo {author} {\bibfnamefont {D.}~\bibnamefont
  {{Pertot}}}, \bibinfo {author} {\bibfnamefont {E.}~\bibnamefont {{Vogt}}},
  \bibinfo {author} {\bibfnamefont {B.}~\bibnamefont {{Fr{\"o}hlich}}},
  \bibinfo {author} {\bibfnamefont {M.}~\bibnamefont {{Feld}}}, \ and\ \bibinfo
  {author} {\bibfnamefont {M.}~\bibnamefont {{K{\"o}hl}}},\ }\bibfield  {title}
  {\bibinfo {title} {\emph {{Attractive and repulsive Fermi polarons in two
  dimensions}}},\ }\href {\doibase 10.1038/nature11151} {\bibfield  {journal}
  {\bibinfo  {journal} {\nat}\ }\textbf {\bibinfo {volume} {485}},\ \bibinfo
  {pages} {619} (\bibinfo {year} {2012})}\BibitemShut {NoStop}%
\bibitem [{\citenamefont {Massignan}\ \emph {et~al.}(2014)\citenamefont
  {Massignan}, \citenamefont {Zaccanti},\ and\ \citenamefont
  {Bruun}}]{Massignan_Zaccanti_Bruun}%
  \BibitemOpen
  \bibfield  {author} {\bibinfo {author} {\bibfnamefont {P.}~\bibnamefont
  {Massignan}}, \bibinfo {author} {\bibfnamefont {M.}~\bibnamefont {Zaccanti}},
  \ and\ \bibinfo {author} {\bibfnamefont {G.~M.}\ \bibnamefont {Bruun}},\
  }\bibfield  {title} {\bibinfo {title} {\emph {Polarons, dressed molecules and
  itinerant ferromagnetism in ultracold Fermi gases}},\ }\href
  {http://stacks.iop.org/0034-4885/77/i=3/a=034401} {\bibfield  {journal}
  {\bibinfo  {journal} {Reports on Progress in Physics}\ }\textbf {\bibinfo
  {volume} {77}},\ \bibinfo {pages} {034401} (\bibinfo {year}
  {2014})}\BibitemShut {NoStop}%
\bibitem [{\citenamefont {Wu}\ \emph {et~al.}(2012)\citenamefont {Wu},
  \citenamefont {Park}, \citenamefont {Ahmadi}, \citenamefont {Will},\ and\
  \citenamefont {Zwierlein}}]{Wu2012}%
  \BibitemOpen
  \bibfield  {author} {\bibinfo {author} {\bibfnamefont {C.-H.}\ \bibnamefont
  {Wu}}, \bibinfo {author} {\bibfnamefont {J.~W.}\ \bibnamefont {Park}},
  \bibinfo {author} {\bibfnamefont {P.}~\bibnamefont {Ahmadi}}, \bibinfo
  {author} {\bibfnamefont {S.}~\bibnamefont {Will}}, \ and\ \bibinfo {author}
  {\bibfnamefont {M.~W.}\ \bibnamefont {Zwierlein}},\ }\bibfield  {title}
  {\bibinfo {title} {\emph {Ultracold Fermionic Feshbach Molecules of
  $^{23}\mathrm{Na}^{40}\mathbf{K}$}},\ }\href {\doibase
  10.1103/PhysRevLett.109.085301} {\bibfield  {journal} {\bibinfo  {journal}
  {Phys. Rev. Lett.}\ }\textbf {\bibinfo {volume} {109}},\ \bibinfo {pages}
  {085301} (\bibinfo {year} {2012})}\BibitemShut {NoStop}%
\bibitem [{\citenamefont {Heo}\ \emph {et~al.}(2012)\citenamefont {Heo},
  \citenamefont {Wang}, \citenamefont {Christensen}, \citenamefont {Rvachov},
  \citenamefont {Cotta}, \citenamefont {Choi}, \citenamefont {Lee},\ and\
  \citenamefont {Ketterle}}]{Heo2012}%
  \BibitemOpen
  \bibfield  {author} {\bibinfo {author} {\bibfnamefont {M.-S.}\ \bibnamefont
  {Heo}}, \bibinfo {author} {\bibfnamefont {T.~T.}\ \bibnamefont {Wang}},
  \bibinfo {author} {\bibfnamefont {C.~A.}\ \bibnamefont {Christensen}},
  \bibinfo {author} {\bibfnamefont {T.~M.}\ \bibnamefont {Rvachov}}, \bibinfo
  {author} {\bibfnamefont {D.~A.}\ \bibnamefont {Cotta}}, \bibinfo {author}
  {\bibfnamefont {J.-H.}\ \bibnamefont {Choi}}, \bibinfo {author}
  {\bibfnamefont {Y.-R.}\ \bibnamefont {Lee}}, \ and\ \bibinfo {author}
  {\bibfnamefont {W.}~\bibnamefont {Ketterle}},\ }\bibfield  {title} {\bibinfo
  {title} {\emph {Formation of ultracold fermionic NaLi Feshbach molecules}},\
  }\href {\doibase 10.1103/PhysRevA.86.021602} {\bibfield  {journal} {\bibinfo
  {journal} {Phys. Rev. A}\ }\textbf {\bibinfo {volume} {86}},\ \bibinfo
  {pages} {021602} (\bibinfo {year} {2012})}\BibitemShut {NoStop}%
\bibitem [{\citenamefont {Cumby}\ \emph {et~al.}(2013)\citenamefont {Cumby},
  \citenamefont {Shewmon}, \citenamefont {Hu}, \citenamefont {Perreault},\ and\
  \citenamefont {Jin}}]{Cumby2013}%
  \BibitemOpen
  \bibfield  {author} {\bibinfo {author} {\bibfnamefont {T.~D.}\ \bibnamefont
  {Cumby}}, \bibinfo {author} {\bibfnamefont {R.~A.}\ \bibnamefont {Shewmon}},
  \bibinfo {author} {\bibfnamefont {M.-G.}\ \bibnamefont {Hu}}, \bibinfo
  {author} {\bibfnamefont {J.~D.}\ \bibnamefont {Perreault}}, \ and\ \bibinfo
  {author} {\bibfnamefont {D.~S.}\ \bibnamefont {Jin}},\ }\bibfield  {title}
  {\bibinfo {title} {\emph {Feshbach-molecule formation in a Bose-Fermi
  mixture}},\ }\href {\doibase 10.1103/PhysRevA.87.012703} {\bibfield
  {journal} {\bibinfo  {journal} {Phys. Rev. A}\ }\textbf {\bibinfo {volume}
  {87}},\ \bibinfo {pages} {012703} (\bibinfo {year} {2013})}\BibitemShut
  {NoStop}%
\bibitem [{\citenamefont {Roati}\ \emph {et~al.}(2007)\citenamefont {Roati},
  \citenamefont {Zaccanti}, \citenamefont {D'Errico}, \citenamefont {Catani},
  \citenamefont {Modugno}, \citenamefont {Simoni}, \citenamefont {Inguscio},\
  and\ \citenamefont {Modugno}}]{Roati2007}%
  \BibitemOpen
  \bibfield  {author} {\bibinfo {author} {\bibfnamefont {G.}~\bibnamefont
  {Roati}}, \bibinfo {author} {\bibfnamefont {M.}~\bibnamefont {Zaccanti}},
  \bibinfo {author} {\bibfnamefont {C.}~\bibnamefont {D'Errico}}, \bibinfo
  {author} {\bibfnamefont {J.}~\bibnamefont {Catani}}, \bibinfo {author}
  {\bibfnamefont {M.}~\bibnamefont {Modugno}}, \bibinfo {author} {\bibfnamefont
  {A.}~\bibnamefont {Simoni}}, \bibinfo {author} {\bibfnamefont
  {M.}~\bibnamefont {Inguscio}}, \ and\ \bibinfo {author} {\bibfnamefont
  {G.}~\bibnamefont {Modugno}},\ }\bibfield  {title} {\bibinfo {title} {\emph
  {$^{39}\mathrm{K}$ Bose-Einstein Condensate with Tunable Interactions}},\
  }\href {\doibase 10.1103/PhysRevLett.99.010403} {\bibfield  {journal}
  {\bibinfo  {journal} {Phys. Rev. Lett.}\ }\textbf {\bibinfo {volume} {99}},\
  \bibinfo {pages} {010403} (\bibinfo {year} {2007})}\BibitemShut {NoStop}%
\bibitem [{\citenamefont {Pilch}\ \emph {et~al.}(2009)\citenamefont {Pilch},
  \citenamefont {Lange}, \citenamefont {Prantner}, \citenamefont {Kerner},
  \citenamefont {Ferlaino}, \citenamefont {N\"agerl},\ and\ \citenamefont
  {Grimm}}]{Pilch2009}%
  \BibitemOpen
  \bibfield  {author} {\bibinfo {author} {\bibfnamefont {K.}~\bibnamefont
  {Pilch}}, \bibinfo {author} {\bibfnamefont {A.~D.}\ \bibnamefont {Lange}},
  \bibinfo {author} {\bibfnamefont {A.}~\bibnamefont {Prantner}}, \bibinfo
  {author} {\bibfnamefont {G.}~\bibnamefont {Kerner}}, \bibinfo {author}
  {\bibfnamefont {F.}~\bibnamefont {Ferlaino}}, \bibinfo {author}
  {\bibfnamefont {H.-C.}\ \bibnamefont {N\"agerl}}, \ and\ \bibinfo {author}
  {\bibfnamefont {R.}~\bibnamefont {Grimm}},\ }\bibfield  {title} {\bibinfo
  {title} {\emph {Observation of interspecies Feshbach resonances in an
  ultracold Rb-Cs mixture}},\ }\href {\doibase 10.1103/PhysRevA.79.042718}
  {\bibfield  {journal} {\bibinfo  {journal} {Phys. Rev. A}\ }\textbf {\bibinfo
  {volume} {79}},\ \bibinfo {pages} {042718} (\bibinfo {year}
  {2009})}\BibitemShut {NoStop}%
\bibitem [{\citenamefont {Combescot}\ and\ \citenamefont
  {Giraud}(2008)}]{Combescot2008}%
  \BibitemOpen
  \bibfield  {author} {\bibinfo {author} {\bibfnamefont {R.}~\bibnamefont
  {Combescot}}\ and\ \bibinfo {author} {\bibfnamefont {S.}~\bibnamefont
  {Giraud}},\ }\bibfield  {title} {\bibinfo {title} {\emph {Normal State of
  Highly Polarized Fermi Gases: Full Many-Body Treatment}},\ }\href {\doibase
  10.1103/PhysRevLett.101.050404} {\bibfield  {journal} {\bibinfo  {journal}
  {Phys. Rev. Lett.}\ }\textbf {\bibinfo {volume} {101}},\ \bibinfo {pages}
  {050404} (\bibinfo {year} {2008})}\BibitemShut {NoStop}%
\bibitem [{\citenamefont {Vlietinck}\ \emph {et~al.}(2013)\citenamefont
  {Vlietinck}, \citenamefont {Ryckebusch},\ and\ \citenamefont
  {Van~Houcke}}]{Vlietinck2013}%
  \BibitemOpen
  \bibfield  {author} {\bibinfo {author} {\bibfnamefont {J.}~\bibnamefont
  {Vlietinck}}, \bibinfo {author} {\bibfnamefont {J.}~\bibnamefont
  {Ryckebusch}}, \ and\ \bibinfo {author} {\bibfnamefont {K.}~\bibnamefont
  {Van~Houcke}},\ }\bibfield  {title} {\bibinfo {title} {\emph {Quasiparticle
  properties of an impurity in a Fermi gas}},\ }\href {\doibase
  10.1103/PhysRevB.87.115133} {\bibfield  {journal} {\bibinfo  {journal} {Phys.
  Rev. B}\ }\textbf {\bibinfo {volume} {87}},\ \bibinfo {pages} {115133}
  (\bibinfo {year} {2013})}\BibitemShut {NoStop}%
\bibitem [{\citenamefont {Mathy}\ \emph {et~al.}(2011)\citenamefont {Mathy},
  \citenamefont {Parish},\ and\ \citenamefont {Huse}}]{Mathy2011}%
  \BibitemOpen
  \bibfield  {author} {\bibinfo {author} {\bibfnamefont {C.~J.~M.}\
  \bibnamefont {Mathy}}, \bibinfo {author} {\bibfnamefont {M.~M.}\ \bibnamefont
  {Parish}}, \ and\ \bibinfo {author} {\bibfnamefont {D.~A.}\ \bibnamefont
  {Huse}},\ }\bibfield  {title} {\bibinfo {title} {\emph {Trimers, molecules
  and polarons in imbalanced atomic {Fermi} gases}},\ }\href@noop {} {\bibfield
   {journal} {\bibinfo  {journal} {Phys. Rev. Lett}\ }\textbf {\bibinfo
  {volume} {106}},\ \bibinfo {pages} {166404} (\bibinfo {year}
  {2011})}\BibitemShut {NoStop}%
\bibitem [{\citenamefont {Orso}\ \emph {et~al.}(2010)\citenamefont {Orso},
  \citenamefont {Burovski},\ and\ \citenamefont {Jolicoeur}}]{orso2010}%
  \BibitemOpen
  \bibfield  {author} {\bibinfo {author} {\bibfnamefont {G.}~\bibnamefont
  {Orso}}, \bibinfo {author} {\bibfnamefont {E.}~\bibnamefont {Burovski}}, \
  and\ \bibinfo {author} {\bibfnamefont {T.}~\bibnamefont {Jolicoeur}},\
  }\bibfield  {title} {\bibinfo {title} {\emph {Luttinger Liquid of Trimers in
  {Fermi} Gases with Unequal Masses}},\ }\href@noop {} {\bibfield  {journal}
  {\bibinfo  {journal} {Phys. Rev. Lett.}\ }\textbf {\bibinfo {volume} {104}},\
  \bibinfo {pages} {065301} (\bibinfo {year} {2010})}\BibitemShut {NoStop}%
\bibitem [{\citenamefont {Parish}(2011)}]{Parish2011}%
  \BibitemOpen
  \bibfield  {author} {\bibinfo {author} {\bibfnamefont {M.~M.}\ \bibnamefont
  {Parish}},\ }\bibfield  {title} {\bibinfo {title} {\emph {Polaron-molecule
  transitions in a two-dimensional Fermi gas}},\ }\href@noop {} {\bibfield
  {journal} {\bibinfo  {journal} {Phys. Rev. A}\ }\textbf {\bibinfo {volume}
  {83}},\ \bibinfo {pages} {051603} (\bibinfo {year} {2011})}\BibitemShut
  {NoStop}%
\bibitem [{\citenamefont {Parish}\ and\ \citenamefont
  {Levinsen}(2013)}]{Parish2013}%
  \BibitemOpen
  \bibfield  {author} {\bibinfo {author} {\bibfnamefont {M.~M.}\ \bibnamefont
  {Parish}}\ and\ \bibinfo {author} {\bibfnamefont {J.}~\bibnamefont
  {Levinsen}},\ }\bibfield  {title} {\bibinfo {title} {\emph {Highly polarized
  {Fermi} gases in two dimensions}},\ }\href@noop {} {\bibfield  {journal}
  {\bibinfo  {journal} {Phys. Rev. A}\ }\textbf {\bibinfo {volume} {87}},\
  \bibinfo {pages} {033616} (\bibinfo {year} {2013})}\BibitemShut {NoStop}%
\bibitem [{\citenamefont {Nishida}(2015)}]{Nishida2015}%
  \BibitemOpen
  \bibfield  {author} {\bibinfo {author} {\bibfnamefont {Y.}~\bibnamefont
  {Nishida}},\ }\bibfield  {title} {\bibinfo {title} {\emph {Polaronic
  Atom-Trimer Continuity in Three-Component Fermi Gases}},\ }\href@noop {}
  {\bibfield  {journal} {\bibinfo  {journal} {Phys. Rev. Lett.}\ }\textbf
  {\bibinfo {volume} {114}},\ \bibinfo {pages} {115302} (\bibinfo {year}
  {2015})}\BibitemShut {NoStop}%
\bibitem [{\citenamefont {Yi}\ and\ \citenamefont {Cui}(2015)}]{cui2015}%
  \BibitemOpen
  \bibfield  {author} {\bibinfo {author} {\bibfnamefont {W.}~\bibnamefont
  {Yi}}\ and\ \bibinfo {author} {\bibfnamefont {X.}~\bibnamefont {Cui}},\
  }\bibfield  {title} {\bibinfo {title} {\emph {Polarons in ultracold Fermi
  superfluids}},\ }\href {\doibase 10.1103/PhysRevA.92.013620} {\bibfield
  {journal} {\bibinfo  {journal} {Phys. Rev. A}\ }\textbf {\bibinfo {volume}
  {92}},\ \bibinfo {pages} {013620} (\bibinfo {year} {2015})}\BibitemShut
  {NoStop}%
\bibitem [{\citenamefont {Astrakharchik}\ and\ \citenamefont
  {Pitaevskii}(2004)}]{Astrakharchik2004}%
  \BibitemOpen
  \bibfield  {author} {\bibinfo {author} {\bibfnamefont {G.~E.}\ \bibnamefont
  {Astrakharchik}}\ and\ \bibinfo {author} {\bibfnamefont {L.~P.}\ \bibnamefont
  {Pitaevskii}},\ }\bibfield  {title} {\bibinfo {title} {\emph {Motion of a
  heavy impurity through a Bose-Einstein condensate}},\ }\href {\doibase
  10.1103/PhysRevA.70.013608} {\bibfield  {journal} {\bibinfo  {journal} {Phys.
  Rev. A}\ }\textbf {\bibinfo {volume} {70}},\ \bibinfo {pages} {013608}
  (\bibinfo {year} {2004})}\BibitemShut {NoStop}%
\bibitem [{\citenamefont {Cucchietti}\ and\ \citenamefont
  {Timmermans}(2006)}]{Cucchietti2006}%
  \BibitemOpen
  \bibfield  {author} {\bibinfo {author} {\bibfnamefont {F.~M.}\ \bibnamefont
  {Cucchietti}}\ and\ \bibinfo {author} {\bibfnamefont {E.}~\bibnamefont
  {Timmermans}},\ }\bibfield  {title} {\bibinfo {title} {\emph {Strong-Coupling
  Polarons in Dilute Gas Bose-Einstein Condensates}},\ }\href {\doibase
  10.1103/PhysRevLett.96.210401} {\bibfield  {journal} {\bibinfo  {journal}
  {Phys. Rev. Lett.}\ }\textbf {\bibinfo {volume} {96}},\ \bibinfo {pages}
  {210401} (\bibinfo {year} {2006})}\BibitemShut {NoStop}%
\bibitem [{\citenamefont {Kalas}\ and\ \citenamefont
  {Blume}(2006)}]{Kalas2006}%
  \BibitemOpen
  \bibfield  {author} {\bibinfo {author} {\bibfnamefont {R.~M.}\ \bibnamefont
  {Kalas}}\ and\ \bibinfo {author} {\bibfnamefont {D.}~\bibnamefont {Blume}},\
  }\bibfield  {title} {\bibinfo {title} {\emph {Interaction-induced
  localization of an impurity in a trapped Bose-Einstein condensate}},\ }\href
  {\doibase 10.1103/PhysRevA.73.043608} {\bibfield  {journal} {\bibinfo
  {journal} {Phys. Rev. A}\ }\textbf {\bibinfo {volume} {73}},\ \bibinfo
  {pages} {043608} (\bibinfo {year} {2006})}\BibitemShut {NoStop}%
\bibitem [{\citenamefont {Bruderer}\ \emph {et~al.}(2008)\citenamefont
  {Bruderer}, \citenamefont {Bao},\ and\ \citenamefont
  {Jaksch}}]{Bruderer2008}%
  \BibitemOpen
  \bibfield  {author} {\bibinfo {author} {\bibfnamefont {M.}~\bibnamefont
  {Bruderer}}, \bibinfo {author} {\bibfnamefont {W.}~\bibnamefont {Bao}}, \
  and\ \bibinfo {author} {\bibfnamefont {D.}~\bibnamefont {Jaksch}},\
  }\bibfield  {title} {\bibinfo {title} {\emph {Self-trapping of impurities in
  Bose-Einstein condensates: Strong attractive and repulsive coupling}},\
  }\href {http://stacks.iop.org/0295-5075/82/i=3/a=30004} {\bibfield  {journal}
  {\bibinfo  {journal} {EPL (Europhysics Letters)}\ }\textbf {\bibinfo {volume}
  {82}},\ \bibinfo {pages} {30004} (\bibinfo {year} {2008})}\BibitemShut
  {NoStop}%
\bibitem [{\citenamefont {Volosniev}\ \emph {et~al.}(2015)\citenamefont
  {Volosniev}, \citenamefont {Hammer},\ and\ \citenamefont
  {Zinner}}]{Volosniev2015}%
  \BibitemOpen
  \bibfield  {author} {\bibinfo {author} {\bibfnamefont {A.~G.}\ \bibnamefont
  {Volosniev}}, \bibinfo {author} {\bibfnamefont {H.-W.}\ \bibnamefont
  {Hammer}}, \ and\ \bibinfo {author} {\bibfnamefont {N.~T.}\ \bibnamefont
  {Zinner}},\ }\bibfield  {title} {\bibinfo {title} {\emph {Real-time dynamics
  of an impurity in an ideal Bose gas in a trap}},\ }\href {\doibase
  10.1103/PhysRevA.92.023623} {\bibfield  {journal} {\bibinfo  {journal} {Phys.
  Rev. A}\ }\textbf {\bibinfo {volume} {92}},\ \bibinfo {pages} {023623}
  (\bibinfo {year} {2015})}\BibitemShut {NoStop}%
\bibitem [{\citenamefont {Huang}\ and\ \citenamefont {Wan}(2009)}]{Huang2009}%
  \BibitemOpen
  \bibfield  {author} {\bibinfo {author} {\bibfnamefont {B.-B.}\ \bibnamefont
  {Huang}}\ and\ \bibinfo {author} {\bibfnamefont {S.-L.}\ \bibnamefont
  {Wan}},\ }\bibfield  {title} {\bibinfo {title} {\emph {Polaron in
  Bose--Einstein--Condensation System}},\ }\href@noop {} {\bibfield  {journal}
  {\bibinfo  {journal} {Chinese Physics Letters}\ }\textbf {\bibinfo {volume}
  {26}},\ \bibinfo {pages} {080302} (\bibinfo {year} {2009})}\BibitemShut
  {NoStop}%
\bibitem [{\citenamefont {Tempere}\ \emph {et~al.}(2009)\citenamefont
  {Tempere}, \citenamefont {Casteels}, \citenamefont {Oberthaler},
  \citenamefont {Knoop}, \citenamefont {Timmermans},\ and\ \citenamefont
  {Devreese}}]{Tempere2009}%
  \BibitemOpen
  \bibfield  {author} {\bibinfo {author} {\bibfnamefont {J.}~\bibnamefont
  {Tempere}}, \bibinfo {author} {\bibfnamefont {W.}~\bibnamefont {Casteels}},
  \bibinfo {author} {\bibfnamefont {M.~K.}\ \bibnamefont {Oberthaler}},
  \bibinfo {author} {\bibfnamefont {S.}~\bibnamefont {Knoop}}, \bibinfo
  {author} {\bibfnamefont {E.}~\bibnamefont {Timmermans}}, \ and\ \bibinfo
  {author} {\bibfnamefont {J.~T.}\ \bibnamefont {Devreese}},\ }\bibfield
  {title} {\bibinfo {title} {\emph {Feynman path-integral treatment of the
  BEC-impurity polaron}},\ }\href {\doibase 10.1103/PhysRevB.80.184504}
  {\bibfield  {journal} {\bibinfo  {journal} {Phys. Rev. B}\ }\textbf {\bibinfo
  {volume} {80}},\ \bibinfo {pages} {184504} (\bibinfo {year}
  {2009})}\BibitemShut {NoStop}%
\bibitem [{\citenamefont {Casteels}\ and\ \citenamefont
  {Wouters}(2014)}]{Casteels2014}%
  \BibitemOpen
  \bibfield  {author} {\bibinfo {author} {\bibfnamefont {W.}~\bibnamefont
  {Casteels}}\ and\ \bibinfo {author} {\bibfnamefont {M.}~\bibnamefont
  {Wouters}},\ }\bibfield  {title} {\bibinfo {title} {\emph {Polaron formation
  in the vicinity of a narrow Feshbach resonance}},\ }\href {\doibase
  10.1103/PhysRevA.90.043602} {\bibfield  {journal} {\bibinfo  {journal} {Phys.
  Rev. A}\ }\textbf {\bibinfo {volume} {90}},\ \bibinfo {pages} {043602}
  (\bibinfo {year} {2014})}\BibitemShut {NoStop}%
\bibitem [{\citenamefont {Shashi}\ \emph {et~al.}(2014)\citenamefont {Shashi},
  \citenamefont {Grusdt}, \citenamefont {Abanin},\ and\ \citenamefont
  {Demler}}]{Shashi2014}%
  \BibitemOpen
  \bibfield  {author} {\bibinfo {author} {\bibfnamefont {A.}~\bibnamefont
  {Shashi}}, \bibinfo {author} {\bibfnamefont {F.}~\bibnamefont {Grusdt}},
  \bibinfo {author} {\bibfnamefont {D.~A.}\ \bibnamefont {Abanin}}, \ and\
  \bibinfo {author} {\bibfnamefont {E.}~\bibnamefont {Demler}},\ }\bibfield
  {title} {\bibinfo {title} {\emph {Radio-frequency spectroscopy of polarons in
  ultracold Bose gases}},\ }\href {\doibase 10.1103/PhysRevA.89.053617}
  {\bibfield  {journal} {\bibinfo  {journal} {Phys. Rev. A}\ }\textbf {\bibinfo
  {volume} {89}},\ \bibinfo {pages} {053617} (\bibinfo {year}
  {2014})}\BibitemShut {NoStop}%
\bibitem [{\citenamefont {{Grusdt}}\ \emph {et~al.}(2015)\citenamefont
  {{Grusdt}}, \citenamefont {{Shchadilova}}, \citenamefont {{Rubtsov}},\ and\
  \citenamefont {{Demler}}}]{Grusdt2014}%
  \BibitemOpen
  \bibfield  {author} {\bibinfo {author} {\bibfnamefont {F.}~\bibnamefont
  {{Grusdt}}}, \bibinfo {author} {\bibfnamefont {Y.~E.}\ \bibnamefont
  {{Shchadilova}}}, \bibinfo {author} {\bibfnamefont {A.~N.}\ \bibnamefont
  {{Rubtsov}}}, \ and\ \bibinfo {author} {\bibfnamefont {E.}~\bibnamefont
  {{Demler}}},\ }\bibfield  {title} {\bibinfo {title} {\emph {{Renormalization
  group approach to the Fr\"ohlich polaron model: application to impurity-BEC
  problem}}},\ }\href@noop {} {\bibfield  {journal} {\bibinfo  {journal} {Sci.
  Rep.}\ }\textbf {\bibinfo {volume} {5}},\ \bibinfo {pages} {12124} (\bibinfo
  {year} {2015})}\BibitemShut {NoStop}%
\bibitem [{\citenamefont {{Shchadilova}}\ \emph {et~al.}(2014)\citenamefont
  {{Shchadilova}}, \citenamefont {{Grusdt}}, \citenamefont {{Rubtsov}},\ and\
  \citenamefont {{Demler}}}]{Shchadilova2014}%
  \BibitemOpen
  \bibfield  {author} {\bibinfo {author} {\bibfnamefont {Y.~E.}\ \bibnamefont
  {{Shchadilova}}}, \bibinfo {author} {\bibfnamefont {F.}~\bibnamefont
  {{Grusdt}}}, \bibinfo {author} {\bibfnamefont {A.~N.}\ \bibnamefont
  {{Rubtsov}}}, \ and\ \bibinfo {author} {\bibfnamefont {E.}~\bibnamefont
  {{Demler}}},\ }\bibfield  {title} {\bibinfo {title} {\emph {{Polaronic mass
  renormalization of impurities in BEC: correlated Gaussian wavefunction
  approach}}},\ }\href@noop {} {\bibfield  {journal} {\bibinfo  {journal}
  {ArXiv e-prints}\ } (\bibinfo {year} {2014})},\ \Eprint
  {http://arxiv.org/abs/1410.5691} {arXiv:1410.5691 [cond-mat.quant-gas]}
  \BibitemShut {NoStop}%
\bibitem [{\citenamefont {Vlietinck}\ \emph {et~al.}(2015)\citenamefont
  {Vlietinck}, \citenamefont {Casteels}, \citenamefont {Houcke}, \citenamefont
  {Tempere}, \citenamefont {Ryckebusch},\ and\ \citenamefont
  {Devreese}}]{Vlietinck2014}%
  \BibitemOpen
  \bibfield  {author} {\bibinfo {author} {\bibfnamefont {J.}~\bibnamefont
  {Vlietinck}}, \bibinfo {author} {\bibfnamefont {W.}~\bibnamefont {Casteels}},
  \bibinfo {author} {\bibfnamefont {K.~V.}\ \bibnamefont {Houcke}}, \bibinfo
  {author} {\bibfnamefont {J.}~\bibnamefont {Tempere}}, \bibinfo {author}
  {\bibfnamefont {J.}~\bibnamefont {Ryckebusch}}, \ and\ \bibinfo {author}
  {\bibfnamefont {J.~T.}\ \bibnamefont {Devreese}},\ }\bibfield  {title}
  {\bibinfo {title} {\emph {Diagrammatic Monte Carlo study of the acoustic and
  the Bose–Einstein condensate polaron}},\ }\href
  {http://stacks.iop.org/1367-2630/17/i=3/a=033023} {\bibfield  {journal}
  {\bibinfo  {journal} {New Journal of Physics}\ }\textbf {\bibinfo {volume}
  {17}},\ \bibinfo {pages} {033023} (\bibinfo {year} {2015})}\BibitemShut
  {NoStop}%
\bibitem [{\citenamefont {Li}\ and\ \citenamefont {Das~Sarma}(2014)}]{Li2014}%
  \BibitemOpen
  \bibfield  {author} {\bibinfo {author} {\bibfnamefont {W.}~\bibnamefont
  {Li}}\ and\ \bibinfo {author} {\bibfnamefont {S.}~\bibnamefont {Das~Sarma}},\
  }\bibfield  {title} {\bibinfo {title} {\emph {Variational study of polarons
  in Bose-Einstein condensates}},\ }\href {\doibase 10.1103/PhysRevA.90.013618}
  {\bibfield  {journal} {\bibinfo  {journal} {Phys. Rev. A}\ }\textbf {\bibinfo
  {volume} {90}},\ \bibinfo {pages} {013618} (\bibinfo {year}
  {2014})}\BibitemShut {NoStop}%
\bibitem [{\citenamefont {Rath}\ and\ \citenamefont
  {Schmidt}(2013)}]{Rath2013}%
  \BibitemOpen
  \bibfield  {author} {\bibinfo {author} {\bibfnamefont {S.~P.}\ \bibnamefont
  {Rath}}\ and\ \bibinfo {author} {\bibfnamefont {R.}~\bibnamefont {Schmidt}},\
  }\bibfield  {title} {\bibinfo {title} {\emph {Field-theoretical study of the
  Bose polaron}},\ }\href {\doibase 10.1103/PhysRevA.88.053632} {\bibfield
  {journal} {\bibinfo  {journal} {Phys. Rev. A}\ }\textbf {\bibinfo {volume}
  {88}},\ \bibinfo {pages} {053632} (\bibinfo {year} {2013})}\BibitemShut
  {NoStop}%
\bibitem [{\citenamefont {{S{\o}gaard Christensen}}\ \emph
  {et~al.}(2015)\citenamefont {{S{\o}gaard Christensen}}, \citenamefont
  {{Levinsen}},\ and\ \citenamefont {{Bruun}}}]{Christensen2015}%
  \BibitemOpen
  \bibfield  {author} {\bibinfo {author} {\bibfnamefont {R.}~\bibnamefont
  {{S{\o}gaard Christensen}}}, \bibinfo {author} {\bibfnamefont
  {J.}~\bibnamefont {{Levinsen}}}, \ and\ \bibinfo {author} {\bibfnamefont
  {G.~M.}\ \bibnamefont {{Bruun}}},\ }\bibfield  {title} {\bibinfo {title}
  {\emph {{Quasiparticle properties of a mobile impurity in a Bose-Einstein
  condensate}}},\ }\href@noop {} {\bibfield  {journal} {\bibinfo  {journal}
  {Phys. Rev. Lett.}\ }\textbf {\bibinfo {volume} {115}},\ \bibinfo {pages}
  {160401} (\bibinfo {year} {2015})}\BibitemShut {NoStop}%
\bibitem [{\citenamefont {Berninger}\ \emph {et~al.}(2011)\citenamefont
  {Berninger}, \citenamefont {Zenesini}, \citenamefont {Huang}, \citenamefont
  {Harm}, \citenamefont {N\"agerl}, \citenamefont {Ferlaino}, \citenamefont
  {Grimm}, \citenamefont {Julienne},\ and\ \citenamefont
  {Hutson}}]{Berninger2011}%
  \BibitemOpen
  \bibfield  {author} {\bibinfo {author} {\bibfnamefont {M.}~\bibnamefont
  {Berninger}}, \bibinfo {author} {\bibfnamefont {A.}~\bibnamefont {Zenesini}},
  \bibinfo {author} {\bibfnamefont {B.}~\bibnamefont {Huang}}, \bibinfo
  {author} {\bibfnamefont {W.}~\bibnamefont {Harm}}, \bibinfo {author}
  {\bibfnamefont {H.-C.}\ \bibnamefont {N\"agerl}}, \bibinfo {author}
  {\bibfnamefont {F.}~\bibnamefont {Ferlaino}}, \bibinfo {author}
  {\bibfnamefont {R.}~\bibnamefont {Grimm}}, \bibinfo {author} {\bibfnamefont
  {P.~S.}\ \bibnamefont {Julienne}}, \ and\ \bibinfo {author} {\bibfnamefont
  {J.~M.}\ \bibnamefont {Hutson}},\ }\bibfield  {title} {\bibinfo {title}
  {\emph {Universality of the Three-Body Parameter for {Efimov} States in
  Ultracold Cesium}},\ }\href@noop {} {\bibfield  {journal} {\bibinfo
  {journal} {Phys. Rev. Lett.}\ }\textbf {\bibinfo {volume} {107}},\ \bibinfo
  {pages} {120401} (\bibinfo {year} {2011})}\BibitemShut {NoStop}%
\bibitem [{\citenamefont {Roy}\ \emph {et~al.}(2013)\citenamefont {Roy},
  \citenamefont {Landini}, \citenamefont {Trenkwalder}, \citenamefont
  {Semeghini}, \citenamefont {Spagnolli}, \citenamefont {Simoni}, \citenamefont
  {Fattori}, \citenamefont {Inguscio},\ and\ \citenamefont
  {Modugno}}]{Roy2013}%
  \BibitemOpen
  \bibfield  {author} {\bibinfo {author} {\bibfnamefont {S.}~\bibnamefont
  {Roy}}, \bibinfo {author} {\bibfnamefont {M.}~\bibnamefont {Landini}},
  \bibinfo {author} {\bibfnamefont {A.}~\bibnamefont {Trenkwalder}}, \bibinfo
  {author} {\bibfnamefont {G.}~\bibnamefont {Semeghini}}, \bibinfo {author}
  {\bibfnamefont {G.}~\bibnamefont {Spagnolli}}, \bibinfo {author}
  {\bibfnamefont {A.}~\bibnamefont {Simoni}}, \bibinfo {author} {\bibfnamefont
  {M.}~\bibnamefont {Fattori}}, \bibinfo {author} {\bibfnamefont
  {M.}~\bibnamefont {Inguscio}}, \ and\ \bibinfo {author} {\bibfnamefont
  {G.}~\bibnamefont {Modugno}},\ }\bibfield  {title} {\bibinfo {title} {\emph
  {Test of the Universality of the Three-Body {Efimov} Parameter at Narrow
  {Feshbach} Resonances}},\ }\href@noop {} {\bibfield  {journal} {\bibinfo
  {journal} {Phys. Rev. Lett.}\ }\textbf {\bibinfo {volume} {111}},\ \bibinfo
  {pages} {053202} (\bibinfo {year} {2013})}\BibitemShut {NoStop}%
\bibitem [{\citenamefont {Wang}\ \emph
  {et~al.}(2012{\natexlab{a}})\citenamefont {Wang}, \citenamefont {D'Incao},
  \citenamefont {Esry},\ and\ \citenamefont {Greene}}]{Wang2012}%
  \BibitemOpen
  \bibfield  {author} {\bibinfo {author} {\bibfnamefont {J.}~\bibnamefont
  {Wang}}, \bibinfo {author} {\bibfnamefont {J.~P.}\ \bibnamefont {D'Incao}},
  \bibinfo {author} {\bibfnamefont {B.~D.}\ \bibnamefont {Esry}}, \ and\
  \bibinfo {author} {\bibfnamefont {C.~H.}\ \bibnamefont {Greene}},\ }\bibfield
   {title} {\bibinfo {title} {\emph {Origin of the Three-Body Parameter
  Universality in {Efimov} Physics}},\ }\href@noop {} {\bibfield  {journal}
  {\bibinfo  {journal} {Phys. Rev. Lett.}\ }\textbf {\bibinfo {volume} {108}},\
  \bibinfo {pages} {263001} (\bibinfo {year} {2012}{\natexlab{a}})}\BibitemShut
  {NoStop}%
\bibitem [{\citenamefont {Bruun}\ and\ \citenamefont
  {Pethick}(2004)}]{Bruun2004}%
  \BibitemOpen
  \bibfield  {author} {\bibinfo {author} {\bibfnamefont {G.~M.}\ \bibnamefont
  {Bruun}}\ and\ \bibinfo {author} {\bibfnamefont {C.~J.}\ \bibnamefont
  {Pethick}},\ }\bibfield  {title} {\bibinfo {title} {\emph {Effective Theory
  of Feshbach Resonances and Many-Body Properties of Fermi Gases}},\ }\href
  {\doibase 10.1103/PhysRevLett.92.140404} {\bibfield  {journal} {\bibinfo
  {journal} {Phys. Rev. Lett.}\ }\textbf {\bibinfo {volume} {92}},\ \bibinfo
  {pages} {140404} (\bibinfo {year} {2004})}\BibitemShut {NoStop}%
\bibitem [{sup()}]{supmat}%
  \BibitemOpen
  \href@noop {} {}\bibinfo {note} {See Supplemental Material for the
  impurity-boson scattering analysis, variational equations, residue, and decay
  rate.}\BibitemShut {Stop}%
\bibitem [{\citenamefont {Wang}\ \emph
  {et~al.}(2012{\natexlab{b}})\citenamefont {Wang}, \citenamefont {Wang},
  \citenamefont {D'Incao},\ and\ \citenamefont {Greene}}]{Wang2012_2}%
  \BibitemOpen
  \bibfield  {author} {\bibinfo {author} {\bibfnamefont {Y.}~\bibnamefont
  {Wang}}, \bibinfo {author} {\bibfnamefont {J.}~\bibnamefont {Wang}}, \bibinfo
  {author} {\bibfnamefont {J.~P.}\ \bibnamefont {D'Incao}}, \ and\ \bibinfo
  {author} {\bibfnamefont {C.~H.}\ \bibnamefont {Greene}},\ }\bibfield  {title}
  {\bibinfo {title} {\emph {Universal Three-Body Parameter in Heteronuclear
  Atomic Systems}},\ }\href@noop {} {\bibfield  {journal} {\bibinfo  {journal}
  {Phys. Rev. Lett.}\ }\textbf {\bibinfo {volume} {109}},\ \bibinfo {pages}
  {243201} (\bibinfo {year} {2012}{\natexlab{b}})}\BibitemShut {NoStop}%
\bibitem [{\citenamefont {Petrov}(2004)}]{Petrov2004tbp}%
  \BibitemOpen
  \bibfield  {author} {\bibinfo {author} {\bibfnamefont {D.~S.}\ \bibnamefont
  {Petrov}},\ }\bibfield  {title} {\bibinfo {title} {\emph {Three-boson problem
  near a narrow Feshbach resonance}},\ }\href@noop {} {\bibfield  {journal}
  {\bibinfo  {journal} {Phys. Rev. Lett.}\ }\textbf {\bibinfo {volume} {93}},\
  \bibinfo {pages} {143201} (\bibinfo {year} {2004})}\BibitemShut {NoStop}%
\bibitem [{\citenamefont {Levinsen}\ \emph {et~al.}(2009)\citenamefont
  {Levinsen}, \citenamefont {Tiecke}, \citenamefont {Walraven},\ and\
  \citenamefont {Petrov}}]{Levinsen2009ads}%
  \BibitemOpen
  \bibfield  {author} {\bibinfo {author} {\bibfnamefont {J.}~\bibnamefont
  {Levinsen}}, \bibinfo {author} {\bibfnamefont {T.~G.}\ \bibnamefont
  {Tiecke}}, \bibinfo {author} {\bibfnamefont {J.~T.~M.}\ \bibnamefont
  {Walraven}}, \ and\ \bibinfo {author} {\bibfnamefont {D.~S.}\ \bibnamefont
  {Petrov}},\ }\bibfield  {title} {\bibinfo {title} {\emph {Atom-dimer
  scattering and long-lived trimers in fermionic mixtures}},\ }\href@noop {}
  {\bibfield  {journal} {\bibinfo  {journal} {Phys. Rev. Lett.}\ }\textbf
  {\bibinfo {volume} {103}},\ \bibinfo {pages} {153202} (\bibinfo {year}
  {2009})}\BibitemShut {NoStop}%
\bibitem [{\citenamefont {Petrov}(2013)}]{Petrov2012tfa}%
  \BibitemOpen
  \bibfield  {author} {\bibinfo {author} {\bibfnamefont {D.~S.}\ \bibnamefont
  {Petrov}},\ }\bibfield  {title} {\bibinfo {title} {\emph {The few-atom
  problem}},\ }in\ \href@noop {} {\emph {\bibinfo {booktitle} {Proceedings of
  the Les Houches Summer Schools, Session 94}}},\ \bibinfo {editor} {edited by\
  \bibinfo {editor} {\bibfnamefont {C.}~\bibnamefont {Salomon}}, \bibinfo
  {editor} {\bibfnamefont {G.~V.}\ \bibnamefont {Shlyapnikov}}, \ and\ \bibinfo
  {editor} {\bibfnamefont {L.~F.}\ \bibnamefont {Cugliandolo}}}\ (\bibinfo
  {publisher} {Oxford University Press, Oxford, England},\ \bibinfo {year}
  {2013})\ p.\ \bibinfo {pages} {109},\ \bibinfo {note} {preprint available at
  arXiv:1206.5752}\BibitemShut {NoStop}%
\bibitem [{\citenamefont {{von Stecher}}\ \emph {et~al.}(2009)\citenamefont
  {{von Stecher}}, \citenamefont {{D'Incao}},\ and\ \citenamefont
  {{Greene}}}]{Stecher2008}%
  \BibitemOpen
  \bibfield  {author} {\bibinfo {author} {\bibfnamefont {J.}~\bibnamefont {{von
  Stecher}}}, \bibinfo {author} {\bibfnamefont {J.~P.}\ \bibnamefont
  {{D'Incao}}}, \ and\ \bibinfo {author} {\bibfnamefont {C.~H.}\ \bibnamefont
  {{Greene}}},\ }\bibfield  {title} {\bibinfo {title} {\emph {{Four-body legacy
  of the Efimov effect}}},\ }\href@noop {} {\bibfield  {journal} {\bibinfo
  {journal} {Nature Physics}\ }\textbf {\bibinfo {volume} {5}},\ \bibinfo
  {pages} {417} (\bibinfo {year} {2009})}\BibitemShut {NoStop}%
\bibitem [{\citenamefont {Blume}\ and\ \citenamefont {Yan}(2014)}]{Blume2014}%
  \BibitemOpen
  \bibfield  {author} {\bibinfo {author} {\bibfnamefont {D.}~\bibnamefont
  {Blume}}\ and\ \bibinfo {author} {\bibfnamefont {Y.}~\bibnamefont {Yan}},\
  }\bibfield  {title} {\bibinfo {title} {\emph {Generalized Efimov Scenario for
  Heavy-Light Mixtures}},\ }\href {\doibase 10.1103/PhysRevLett.113.213201}
  {\bibfield  {journal} {\bibinfo  {journal} {Phys. Rev. Lett.}\ }\textbf
  {\bibinfo {volume} {113}},\ \bibinfo {pages} {213201} (\bibinfo {year}
  {2014})}\BibitemShut {NoStop}%
\bibitem [{\citenamefont {Braaten}\ and\ \citenamefont
  {Hammer}(2001)}]{Braaten2001}%
  \BibitemOpen
  \bibfield  {author} {\bibinfo {author} {\bibfnamefont {E.}~\bibnamefont
  {Braaten}}\ and\ \bibinfo {author} {\bibfnamefont {H.-W.}\ \bibnamefont
  {Hammer}},\ }\bibfield  {title} {\bibinfo {title} {\emph {Three-Body
  Recombination into Deep Bound States in a Bose Gas with Large Scattering
  Length}},\ }\href {\doibase 10.1103/PhysRevLett.87.160407} {\bibfield
  {journal} {\bibinfo  {journal} {Phys. Rev. Lett.}\ }\textbf {\bibinfo
  {volume} {87}},\ \bibinfo {pages} {160407} (\bibinfo {year}
  {2001})}\BibitemShut {NoStop}%
\bibitem [{\citenamefont {Levinsen}\ and\ \citenamefont
  {Petrov}(2011)}]{Levinsen2011}%
  \BibitemOpen
  \bibfield  {author} {\bibinfo {author} {\bibfnamefont {J.}~\bibnamefont
  {Levinsen}}\ and\ \bibinfo {author} {\bibfnamefont {D.}~\bibnamefont
  {Petrov}},\ }\bibfield  {title} {\bibinfo {title} {\emph {Atom-dimer and
  dimer-dimer scattering in fermionic mixtures near a narrow Feshbach
  resonance}},\ }\href@noop {} {\bibfield  {journal} {\bibinfo  {journal} {The
  European Physical Journal D}\ }\textbf {\bibinfo {volume} {65}},\ \bibinfo
  {pages} {67} (\bibinfo {year} {2011})}\BibitemShut {NoStop}%
\bibitem [{\citenamefont {Levinsen}\ \emph {et~al.}(2014)\citenamefont
  {Levinsen}, \citenamefont {Massignan},\ and\ \citenamefont
  {Parish}}]{Levinsen2014}%
  \BibitemOpen
  \bibfield  {author} {\bibinfo {author} {\bibfnamefont {J.}~\bibnamefont
  {Levinsen}}, \bibinfo {author} {\bibfnamefont {P.}~\bibnamefont {Massignan}},
  \ and\ \bibinfo {author} {\bibfnamefont {M.~M.}\ \bibnamefont {Parish}},\
  }\bibfield  {title} {\bibinfo {title} {\emph {Efimov Trimers under Strong
  Confinement}},\ }\href@noop {} {\bibfield  {journal} {\bibinfo  {journal}
  {Phys. Rev. X}\ }\textbf {\bibinfo {volume} {4}},\ \bibinfo {pages} {031020}
  (\bibinfo {year} {2014})}\BibitemShut {NoStop}%
\end{thebibliography}%

%\end{document}

\newpage

%%%%%%%%%% Merge with supplemental materials %%%%%%%%%%
\widetext
\clearpage
\begin{center}
\textbf{\large Supplemental Material:
			Impurity in a Bose-Einstein condensate and the Efimov effect}\\
\vspace{4mm}
{Jesper~Levinsen,$^1$ Meera~M.~Parish,$^{1,2}$ and
  Georg~M.~Bruun$^3$}\\
\vspace{2mm}
{\em \small
$^1$School of Physics and Astronomy, Monash University, Victoria 3800, Australia\\
$^2$London Centre for Nanotechnology, Gordon Street, London, WC1H 0AH, United Kingdom\\
$^3$Department of Physics and Astronomy, Aarhus University, DK-8000 Aarhus C, Denmark
}\end{center}
%%%%%%%%%% Merge with supplemental materials %%%%%%%%%%
%%%%%%%%%% Prefix a "S" to all equations, figures, tables and reset the counter %%%%%%%%%%
\setcounter{equation}{0}
\setcounter{figure}{0}
\setcounter{table}{0}
\setcounter{page}{1}
\makeatletter
\renewcommand{\theequation}{S\arabic{equation}}
\renewcommand{\thefigure}{S\arabic{figure}}
%\renewcommand{\bibnumfmt}[1]{[S#1]}
%\renewcommand{\citenumfont}[1]{S#1}
%%%%%%%%%% Prefix a "S" to all equations, figures, tables and reset the counter %%%%%%%%%%

\section{Impurity-boson scattering}
The bare molecule energy $\nu_0$ is not a physical observable since
the molecule is dressed by open channel impurity-boson pairs.
Likewise, the coupling is assumed to be a constant $g$ for momenta below an
ultraviolet cut-off $\Lambda$, which is not related directly to any
observable. We can relate these parameters to observables by considering
impurity-boson scattering via the closed channel molecule. In a
vacuum, the scattering matrix is~\cite{Bruun2004,Levinsen2011}
\begin{align}
{\mathcal T}_{\text{vac}}(\omega)=\frac{g^2}{\omega-\nu_0-g^2\Pi_{\text{vac}}(0)-g^2\Delta\Pi_{\text{vac}}(\omega)}
\label{Tvac}
\end{align}
where $\Pi_{\text{vac}}(\omega)=\sum_{\mathbf k}(\omega-k^2/2m_r)^{-1}$
is the vacuum pair propagator with energy $\omega$, $m_r=m/2$ is the reduced mass, and
$\Delta\Pi_{\text{vac}}(\omega)=\Pi_{\text{vac}}(\omega)-\Pi_{\text{vac}}(0)$. Equating
(\ref{Tvac}) evaluated at the scattering energy $\omega=p^2/2m_r$
with the effective range expansion
${\mathcal T}_{\text{vac}}(p^2/2m_r)=2\pi a/m_r(1-k^2ar_0/2+ika)$,
where $r_0$ is the effective range, we obtain
$a=m_rg^2(g^2m_r\Lambda/\pi^2-\nu_0)/2\pi$ and
$r_0=-2\pi/m_r^2g^2$. In this way we replace the bare parameters
$\nu_0$ and $\Lambda$ with the low energy observables $a$ and $r_0$
for the impurity-boson scattering.
 %Contrast with Frohlich model, where there are issues with UV regularization
Then we are at liberty to take the limits $\nu_0 \to \infty$, $\Lambda \to \infty$.

\section{Variational equations}
Taking the derivative $\bra{\partial\psi} (\hat{H} - E) \ket{\psi}=0$  with respect to $\alpha_0$, $ \alpha_\k $, $\alpha_{\k_1 \k_2}$, $\gamma_0$, and $\gamma_\k$
yields the five equations 
\begin{align}
E \alpha_0 = & g \sqrt{n_0} \gamma_0 - g \sum_\k v_\k \gamma_\k,
                \nn \\
\left(E-\epsilon_\k - E_\k  \right) \alpha_\k 
=  &  g u_\k \gamma_0 +  g\sqrt{n_0} \gamma_\k, \nn \\
E_{\k_1 \k_2} \alpha_{\k_1 \k_2} = &  g \left(\gamma_\vect{k_1}
  u_\vect{k_2} 
+ \gamma_\vect{k_2} u_\vect{k_1}  \right), \nn \\
(E-\nu_0) \gamma_0 = &  g \sqrt{n_0} \alpha_0 + g \sum_\k u_\k
                        \alpha_\k, \nn\\
\left( E- \! \epsilon^{\rm d}_\k \!- \!\nu_0 \!-\! E_\k  \right) \gamma_\k
  = &  
g \sqrt{n_0} \alpha_\k - g  v_\k \alpha_0 + g\sum_{\k'} u_{\k'}
 \alpha_{\k\k'},
\end{align}
where $E_{\k_1\k_2}\equiv E-E_{\k_1}-E_{\k_2}-\epsilon_{\k_1+\k_2}$.

\section{Quasiparticle residue}
Assuming that the variational wavefunction in Eq.~\eqref{eq:psi} is
normalized, $\langle\psi|\psi\rangle=1$, the residue is given by $Z=|\alpha_0|^2$. Carrying out the
renormalization procedure, we arrive at
\begin{align} 
 Z & = \frac{\left(\frac{\sqrt{n_0}\eta}E -  \sum_\k \frac{v_\k \xi_\k}E
     \right)^2}{
\left(\frac{\sqrt{n_0} \eta}E - \sum_\k
\frac{v_\k \xi_\k}E \right)^2 + \frac{\eta^2}{g^2} +
   \sum_\k \left(\frac{\sqrt{n_0}\xi_\k + u_\k
       \eta}{E - \epsilon_\k^{\rm im} - E_\k}\right)^2 
+
   \sum_\k \frac{\xi_\k^2}{g^2} +\sum_{\k_1
     \k_2} \frac{\left(u_{\k_1} \xi_{\k_2} + u_{\k_2}
     \xi_{\k_1} \right)^2}{2E_{\k_1 \k_2}^2}
}.
\end{align}

\section{Decay rate}
In the decay rate, we need to evaluate $\sum_\k |\gamma_\k|^2$. This
is evaluated similarly to $Z$ above:
\begin{align} 
\sum_\k |\gamma_\k|^2 & = \frac{\sum_\k \frac{\xi_\k^2}{g^2}}{
\left(\frac{\sqrt{n_0} \eta}E - \sum_\k
\frac{v_\k \xi_\k}E \right)^2 + \frac{\eta^2}{g^2} +
   \sum_\k \left(\frac{\sqrt{n_0}\xi_\k + u_\k
       \eta}{E - \epsilon_\k^{\rm im} - E_\k}\right)^2 
+
   \sum_\k \frac{\xi_\k^2}{g^2} +\sum_{\k_1
     \k_2} \frac{\left(u_{\k_1} \xi_{\k_2} + u_{\k_2}
     \xi_{\k_1} \right)^2}{2E_{\k_1 \k_2}^2}
}.
\end{align}

\end{document}